\documentclass[reprint,superscriptaddress,amsmath,amssymb,aps,prl,floatfix]{revtex4-2}
\usepackage{siunitx}
\usepackage{makecell}
\usepackage[colorlinks,linkcolor=black,anchorcolor=blue, urlcolor=blue,citecolor=blue,unicode]{hyperref}
\usepackage{multirow}
\usepackage{subfigure}
\usepackage{color}
\usepackage[normalem]{ulem}
\usepackage{amsmath}
\usepackage{booktabs,siunitx,array}
\usepackage{booktabs}
\usepackage{ulem}
\usepackage{listings}
\usepackage{import}
\usepackage{xcolor}
\usepackage{framed}
\usepackage{mdwlist}
\usepackage{diagbox}
\usepackage{bm}
\usepackage{graphicx}
\usepackage{dcolumn}
\usepackage{algorithm}
\usepackage{algpseudocode}
\usepackage{csquotes}
\usepackage{array}
\usepackage{subfigure}
\usepackage{siunitx}
\usepackage{amsmath}
\usepackage{ifthen}
\usepackage{threeparttable}
\usepackage{float}

\usepackage{CJK}
\usepackage{indentfirst}

\begin{document}

\title{\textit{Ab initio}  correlated calculations without finite basis-set error: Numerically precise all-electron RPA correlation energies for diatomic molecules}
\author{Hao Peng}
\affiliation{Institute of Physics, Chinese Academy of Sciences, Beijing 100190, China}
\affiliation{University of Chinese Academy of Sciences, Beijing 100049, China}
\author{Xinguo Ren}
\email{renxg@iphy.ac.cn}
\affiliation{Institute of Physics, Chinese Academy of Sciences, Beijing 100190, China}

\newcommand{\XR}[1]{\color{red}{\bf #1 }}
\newcommand{\XB}[1]{\color{blue}{ #1 }}
\newcommand{\XY}[1]{\color{yellow}{\bf #1 }}

\begin{abstract}
In wavefunction-based \textit{ab-initio} quantum mechanical calculations, achieving absolute convergence with respect to the one-electron basis set is a long-standing challenge. 
In this work, using the
random phase approximation (RPA) electron correlation energy as an example, we show how to compute the basis-error-free RPA correlation energy for
diatomic molecules by iteratively solving the Sternheimer equations for first-order wave functions in the prolate spheroidal coordinate system. 
Our approach provides RPA correlation energies across the periodic table to any desired precision;
in practice, the convergence of the absolute RPA energies to the meV-level accuracy can be readily attained. Our method thus provides unprecedented reference numbers that can be used to assess the
reliability of the commonly used computational procedures in quantum chemistry.
 The numerical techniques
developed in the present work also have direct implications for the development of basis error-free schemes for the GW method or the \textit{ab initio} 
quantum chemistry methods such as MP2.
\end{abstract}

\maketitle

\textit{Introduction.}-
The \textit{ab initio} solution of the many-electron Schr\"{o}dinger equation requires convergence with respect to both the correlated methods and the one-electron basis set. 
The first aspect concerns how much electron correlation is captured in a given Hilbert space, while the second aspect characterizes the completeness of the Hilbert space
itself. Various correlated methods have been developed in quantum chemistry and condensed matter physics
to describe the interacting many-electron systems
\cite{Szabo/Ostlund:1989,Marin/Reining/Ceperley:2016,Friesner:2005,Ghosh/etal:2018},
and considerable effort has been devoted to benchmark their performance in prototypical problems \cite{Eriksen/etal:JPCL2020,Motta/etal:PRX2020}. However,
these benchmarks are usually performed within a small Hilbert space defined typically in terms of a linear combination of atomic orbitals (LCAO). 
Convergence with respect to the one-electron basis sets is usually attained by fitting the results
obtained at finite basis sets in terms of an empirical formula,
attempting to reach the so-called complete basis set (CBS) limit via extrapolation. 
As such, the accuracy of the extrapolated CBS results necessarily depends on the availability and quality of systematically improvable one-electron basis sets and the reliability of the extrapolation rule itself. 
It is therefore highly desirable to find alternative ways to establish the converged limit of correlated methods without relying on finite, atomic basis set. The challenges of developing numerically precise approaches for electronic structure calculations have recently been discussed in Ref.~\cite{lehtola2019review}. While this work addresses the challenge of attaining the CBS results within the LCAO approach, it is important to note that the development of extrapolation schemes within plane-wave implementations of correlated methods is a concurrently active field
\cite{Shepherd/etal:2012,Masios/etal:2024}.

Among various electronic structure methods, the random phase approximation (RPA) stands out as a cornerstone
that bridges quantum chemistry methods and density functional theory. Practical applications show
that RPA-based methods are suitable for describing delicate ground-state energy differences for both molecular and extended systems.  However, a major hurdle in the RPA calculations (and other correlated methods as well) is its very slow convergence with respect to 
one-electron basis set, rendering the numerically fully converged results difficult to obtain. Calculating numerically converged RPA atomization energies via extrapolation is highly non-trivial \cite{humer2022approaching}.  Here, we develop an approach that allows one to obtain numerically precise RPA electron correlation energy without suffering from the basis-set incompleteness error (BSIE). The essential idea behind this approach 
is to compute the non-interacting Kohn-Sham (KS) density response function, the central quantity of RPA, directly from the first-order KS wavefunctions (WFs), which themselves are determined by solving the Sternheimer equation. 
The key point here is that, instead of expanding the Sternheimer equation in terms of a finite basis set, 
we solve it on a real-space grid, arriving at a solution that can be made arbitrary accurate. With a numerically fully converged KS density-response function, one can then obtain the absolute RPA correlation energy without BSIE. This technique has previously been applied to
isolated atoms \cite{peng2023basis}, whereby the 3-dimensional (3D) Sternheimer equation reduces to a 1-dimensional (1D) radial differential equation, which can be conveniently solved on a dense logarithmic grid. In this work, we extend this technique to diatomic molecules by solving the 2-dimensional (2D) radial Sternheimer equation in the prolate spheroidal coordinates system. Previous works have demonstrated that, using prolate spheroidal coordinates, numerically highly precise all-electron ground states of diatomic molecules can be obtained at the level of 
Hartree-Fock and conventional density functional approximations \cite{becke1982numerical,laaksonen1986fully,kobus1996numerical,lehtola2019review}, as well as the
exchange-only optimized effective potential method \cite{makmal2009fully}. This extension enables us to compute the fully converged absolute RPA correlation energy for diatomic molecules. In this context, we mention that there have been continuing efforts over the decades devoted to fully numerical implementations of various electronic structure methods, ranging from  Hartree-Fock
\cite{McCullough:1986,becke1982numerical,laaksonen1986fully,kobus1996numerical}, conventional density functional approximations \cite{lehtola2019review},
the exchange-only optimized effective potential method \cite{makmal2009fully}, as well as MP2 \cite{Artemyev:2004} and coupled cluster (CC) methods \cite{Ludwik/Bartlett:1985}, based on the prolate spheroidal coordinates that are particularly suitable for diatomic molecules. More recently, significant progress has been made in applying multi-wavelet methods to MP2 \cite{Bischoff/Valeev:2013} and CC methods \cite{Kottmann/Bischoff:2017_01,Kottmann/Bischoff:2017_02}, and to solve 3-dimensional Sternheimer equations for calculating static electric polarizability \cite{Brakestad:2020} and magnetic properties \cite{Jensen/etal:2016} for molecules. However, applying the real-space numerical methods to the calculation of RPA correlation energies that can deliver controllable numerical precision, thereby enabling the
calibration of the BSIE in the conventional LCAO approach, has not been reported. 

In the present work, by treating diatomic molecules in a numerically precise way, we can obtain fully converged all-electron (AE) RPA binding energy curves for any diatomic molecules across the periodic table. This permits a rigorous assessment of the BSIE of the commonly used gaussian-type orbital (GTO) basis sets in quantum chemistry, and further quantify the errors arising from the commonly used frozen-core (FC) approximation and the basis set superposition errors (BSSEs). Moreover, such reference data can also guide the development of more efficient atomic basis sets suitable for correlated calculations, especially for extended systems where currently available GTO basis sets have severe deficiencies. Last but not least, this approach can be
directly extended to scaled-opposite-spin second-order M{\o}ller-Plesset perturbation theory (SOS-MP2) and the $GW$ methods, and thus has broad implications for correlated calculations.

\textit{Result.}-We first show that the all-electron RPA correlation energy for diatomic molecules can be converged to arbitrary precision within our approach. Details of our methodology are discussed in the \textit{End Matter} section and the Supplemental Material (SM). Essentially, our approach boils down to iteratively determining the eigenspectum
of the composite $\chi^0(i\omega)v$ operator, where $\chi^0$ is the non-interacting density response function and $v$ the Coulomb interaction. To this end, a key step is to solve the Sternheimer equation of the first-order wavefunction on a prolate spheroidal coordinates system, suitable for diatomic molecules.
The contribution of each eigenstate of the $\chi^0(i\omega)v$ operator to the RPA correlation energy is positively correlated with the magnitude of its eigenvalue. Concretely, the RPA correlation energy for a diatomic molecule is given by
\begin{equation}
    E_c^\mathrm{RPA}
    =\frac{1}{2\pi}\int_0^\infty {{\rm d}\omega}\sum_{M=0}^{M_{\text{max}}}\sum_i^{N_{\text{eigen}}}[ln(1-e_{i,M})+e_{i,M}] \, 
 \label{eq:EcRPA}
 \end{equation}
where $e_{i,M}$ is the $i$-th eigenvalue of $\chi^0(i\omega)v$ in the $M$-th angular momentum channel. Here, the rotational symmetry allows to  
group the eigenvectors of $\chi^0(i\omega)v$  into  different $M$ channels.
In practical calculations, the eigenvalues $e_{i,M}$ can be arranged in the order of their magnitude, from large to small, so that the contribution to the RPA correlation energy will become gradually smaller. Systematically increasing the number of eigenvalues in a given magnetic quantum channel $M$, one can achieve arbitrary precision for a given $M$. Then, keeping increasing the magnetic quantum number $M$ and repeating the above process, one can finally obtain  
the RPA correlation energy to any desired precision. 
According to Eq.~\ref{eq:EcRPA}, the maximum magnetic quantum number $M_{\text{max}}$, and the number of eigenvalues $N_{\text{eigen}}$ for each $M$ are obviously the two key parameters controlling the precision of the calculation. In addition, the number of frequency points and the real-space grid points in the prolate spheroidal coordinates system are two other parameters that affect the numerical precision. The frequency integration is the easiest and the discretization error can be easily reduced to the $\mu$eV level using the modified Gauss-Legendre grid \cite{ren2012resolution} or the minimax grid \cite{Kaltak/Klimes/Kresse:2014}. This has been previously shown in the literature \cite{Azizi/etal:2024} and will not be discussed here. 

Next, we take the N$_2$ molecule (with an {\XB{experimental}} equilibrium bond length of $1.098$ \AA) as a concrete example to check the convergence behavior of the absolute RPA correlation energy with respect to the grid size $(N_\mu, N_\nu)$, as well as $N_{\text{eigen}}$ and $M_{\text{max}}$. 
Regarding the real-space grid $(N_\mu,N_\nu)$, as is demonstrated in Table~S1 of the SM, one can achieve a numerical precision below 1 meV using a grid size of $(90,90)$ and this can be further refined to 0.1 meV when doubling the grid size.  The convergence tests for heavier elements are also provided in the SM (Tables~S2 and S3). We find that numerical precision better than 1 meV can be achieved for all-electron RPA correlation energies for a real-space grid size of the order of $O(100)\times O(100)$. 
In subsequent calculations, we ensure that the grids are dense enough so that the absolute energy is converged within 1 meV.


%

Now we proceed to examine the convergence behavior with respect to the two key parameters: $N_{\text{eigen}}$ and $M_{\text{max}}$. 
 In Table~\ref{tab:1} we present the RPA correlation energies for both the N$_2$ molecule and the isolated N atom
 for increasing number of eigenvalues $N_{\text{eigen}}$, whereby $M_{\text{max}}$ is fixed at 9. All the RPA calculations in the work are performed on top of the Perdew-Burke-Ernzerhof (PBE) \cite{perdew1996generalized} reference orbitals, but the approach works equally well if one starts with other references. For a balanced treatment, in atomic calculations only $N_{\text{eigen}}/2$ eigenvalues are included. The obtained binding energies of the N$_2$ molecule for different $N_{\text{eigen}}$ are also presented in Table~\ref{tab:1}. We see that when $N_{\text{eigen}}$ is increased from 900 to 1000, the absolute RPA correlation energies of N$_2$ and N changes by 0.15 and 0.08 meV, respectively; the binding energy is only changed by 0.001 meV. 
 We note that such a convergence pattern does not change with $M_{\text{max}}$.
\begin{table}[!h]
\caption{RPA@PBE correlation energies (in eV) for N$_{2}$ and the N atom for different numbers of eigenvalues $N_{\text{eigen}}$
with $M_{\text{max}}=9$. Note that only half of $N_{\text{eigen}}$ eigenvalues is used for the N atom calculation.  Here the grid density is (150,120).
The binding energies $\Delta E^\text{RPA}_c = E^\text{RPA}_c(\text{N}_2) - 2E^\text{RPA}_c(\text{N})$) 
are presented in the fourth column.}
\begin{tabular}{c c c c}
\hline
\textbf{$N_{\text{eigen}}$} & $E^\text{RPA}_c(\text{N}_{2})$ &  $E^\text{RPA}_c(\text{N})$ & $\Delta E^\text{RPA}_c$ \\ \hline
100  &   -23.36237    &  -9.22194    & -4.91849 \\ 
200  &   -23.38841    &   -9.23521   & -4.91800 \\
300  &    -23.39446  &     -9.23829 & -4.91789 \\
400  &   -23.39687   &     -9.23951 &  -4.91785\\
500  &   -23.39808   &      -9.24013&  -4.91782\\
600  &    -23.39879  &   -9.24049   & -4.91781 \\
700  &    -23.39923  &      -9.24071& -4.91780 \\
800  &    -23.39954  &     -9.24087  &  -4.91780\\
900  &    -23.39975  &         -9.24098      & -4.91779 \\
1000  &    -23.39990  &    -9.24106  & -4.91778 \\

      \hline
      \hline
     \label{tab:1}
\end{tabular}
\end{table}

\begin{table}[!h]

\caption{Contributions to the RPA@PBE correlation energies of $N_{2}$ and $N$ from different $M$ channels. For each $M$, 
$N_{\text{eigen}}=1000$ and 500 are used for the $N_{2}$ and $N$ calculations, respectively. 
The fourth column represents the $M$-resolved contributions to the RPA correlation part of the binding energy.}
\begin{tabular}{c c c c c}

\hline
\textbf{$M$} & $E^\text{RPA}_c(\text{N}_{2})$ & $E^\text{RPA}_c(\text{N})$   & $\Delta E^\text{RPA}_c$\\ \hline
0  &    -12.32344  &  -4.32484    &    -3.67376 \\
1   &   -8.40979   &  -3.83432     &   -0.74116 \\
2   &   -1.84721   &   -0.71361     &   -0.42000 \\
3    &   -0.51048  &   -0.22959     &   -0.05129 \\ 
4    &   -0.16809  &   -0.07550     &    -0.01709 \\  
5    &  -0.07027 &  -0.03155     &     -0.00717  \\    
6    &  -0.03426 & -0.01537     &    -0.00351  \\     
7    &   -0.01860 &  -0.00834    &     -0.00192   \\    
8    &   -0.01093 &  -0.00489    &    -0.00114  \\
9    &    -0.00682&  -0.00305    &    -0.00072   \\
Total &   -23.39990   & -9.24106  &   -4.91778    \\
      \hline
      \hline
       \label{tab:2}
\end{tabular}
\end{table}

Finally, we examine the convergence behavior of the RPA correlation energy with respect to the maximum magnetic quantum number $M_{\text{max}}$. 
As indicated in Eq.~\ref{eq:EcRPA}, the full RPA correlation energy can be decomposed into contributions from different $M$ channels. 
In Table~\ref{tab:2}, we present the $M$-resolved contributions to the absolute RPA correlation energies of both N$_2$ and N, as well as to their differences (the binding energies), up to $M=9$.  
It should be mentioned that, except for $M=0$, the value for each $M$ in Table~\ref{tab:2}  represents the sum of the results of two degenerate channels $\pm M$. 
From Table~\ref{tab:2}, it can be seen that, despite the rapid decrease in the contribution to $E_c^\text{RPA}$ as $M$ increases, a sizable contribution
of 7 meV is still visible for $M=9$. Obviously, to achieve an absolute convergence of the RPA correlation energy to the meV accuracy, a much larger $M_{\text{max}}$ is needed.

In this connection, we conducted a theoretical analysis of the convergence behavior of the RPA correlation energy of diatomic molecules with respect to $M_{\text{max}}$, and found that the correlation energy should converge as $1/M_{\text{max}}^3$, i.e.,
\begin{equation}
    \label{eq:converge_behaviour}
    E_c^\text{RPA}(M_{\text{max}}) = E_c^\text{RPA}(\infty) + \frac{\alpha}{M_{\text{max}}^3}\, .
\end{equation}
This is similar to the atomic case where the correlation energy converges as $1/L_{\text{max}}^3$ with $L_{\text{max}}$ being the maximum azimuthal quantum
number \cite{Schwartz:1962,Hill:1985,peng2023basis}. 
An in-depth theoretical derivation of this asymptotic behavior and numerical verification
are provided in Sec.~III~B of the SM. Figs.~S2 and S3 in the SM show that the RPA correlation
energies of both the N atom and N$_2$ as a function $M_{\text{max}}$ follow perfectly the behavior given by Eq.~\ref{eq:converge_behaviour} with a conference level (the $R^2$ value) higher than
$99.995\%$. 
Fitting the data from $M_{\text{max}}=10$ to 16, we obtain
$E_c^\text{RPA}(\infty)=-23.41679$ eV for the N$_2$ molecule, with the coefficient $\alpha=12.45267$ eV. Thus, the residual BSIE due to contributions beyond
$M_{\text{max}}$ is $12.45267/M_{\text{max}}^3$ eV, amounting to about 17 meV for $M_{\text{max}}=9$. To have the absolute energy converged to 1 meV, a value of $M_{\text{max}}=24$ is expected. 
A similar fitting for the N-atom results
yields $E_c^\text{RPA}(\infty)=-9.24857$, and $\alpha=5.52887$. Now, since the absolute RPA correlation energies for both N$_2$ and N follow
the same asymptotic behavior, it immediately follows that the binding energies yielded
their differences, also follow the same behavior as Eq.~\ref{eq:converge_behaviour}, although with a much smaller coefficient. 
A fitting for the binding energies yields
$\Delta E_c^\text{RPA}(\infty)= -4.91965$, and a coefficient $\alpha=1.39493$ eV. Therefore, at $M_{\text{max}}=9$, the BSIE for the RPA binding energies
is about 1.9 meV.


%
%


    

Thus, we have completely resolved the convergence problem with respect to $M_{\text{max}}$ for diatomic molecules. In practical calculations, choosing $M_{\text{max}}=9$  can yield binding energies that reach meV-level accuracy. Of course, the BSIE with respect to $M_{\text{max}}$ can be completely eliminated either 
by running calculations with a large $M_{\text{max}}$ (say $M_{\text{max}}=24$) or extrapolating the results to $M_{\text{max}}\rightarrow \infty$ using Eq.~\ref{eq:converge_behaviour}.



Since the absolute AE-RPA correlation energy for diatomic molecules can now be attained
to any desired precision, we naturally have access to very accurate full RPA total energy $E^\text{RPA} = E^\text{HF}\left[\{\psi_i^\text{KS}\}\right]+E^\text{RPA}_c$, where 
the Hartree-Fock part $E^\text{HF}$ of the RPA total energy only involves
occupied states and can be accurately calculated using the standard NAO basis sets as used in the FHI-aims code \cite{blum2009ab}. 
Therefore, we now have a tool to assess the reliability of the computational protocols commonly used
in quantum chemistry, such as the counterpoise (CP) correction to the BSSEs, the FC approximation, and the extrapolation to the CBS limit in terms of empirical fitting formula. In typical correlated
calculations, since it is nearly impossible to converge absolute total energies, the energy differences instead become 
the targeted quantity to converge. However, several factors, such as the choice of basis set, whether or not to perform CP corrections, and the details of the extrapolation procedure, may affect the CBS results obtained, giving rise to remaining uncertainties which are difficult to estimate. For example, 
the RPA interaction energy of the water dimer converges slowly with basis sets, where the CP-corrected and uncorrected CBS limit using Dunning's cc-pV$n$Z (n=5 and 6) basis sets still differ by 4 meV \cite{tahir24}. 

In Fig.~S4 of the SM, we compare the BSSEs present in the traditional sum of states scheme and the present approach for three series of AO basis sets, again using N$_2$ as an illustration example. 
As is evident, in the traditional approach, the BSSEs are huge and do not
vanish even with the largest available basis set in each series. Such BSSEs mainly come from the imbalanced
description of the unoccupied manifold of the Hilbert space of the molecule and the atom. In contrast, in the
present approach, the BSSEs are vanishingly small, as expected. Apparently, in our approach, there is no need to invoke the counterpoise procedure to correct BSSEs any more.

The core-core and core-valence electron correlations usually contribute little to the binding energies, and it is a common practice to employ the FC
approximation in correlated calculations. However, 
the actual error due to the FC approximation is often hard to rigorously assess, especially for heavy elements.  This is because within conventional approaches,
the commonly used atomic orbital basis sets are not completely balanced in describing the valence electrons and core electrons, the BSIE
and FC error are intervened. Our approach here eliminates the BSIEs on equal footings for both AE and FC calculations and thus provides an unambiguous way to assess the FC errors. In Figs.~S5 of the SM, we present the AE and FC binding energy curves for N$_2$, P$_2$, and As$_2$. For P$_2$ and As$_2$, we also freeze different core shells and observe how the results will change. 
The results indicate that freezing all core electrons leads to an increase in errors from N to As
(48 meV for N$_2$, 52 meV for P$_2$, and 160 meV for As$_2$). However, if one just unfreezes the outermost core electrons ($1s$ for N, $2s2p$ for P, and $3s3p3d$ for As), the obtained FC results become almost indistinguishable from the AE results. This investigation clarifies how to perform proper FC calculations without compromising accuracy.

\begin{figure}[htbp]
    
    \centering
    \includegraphics[scale=0.35]{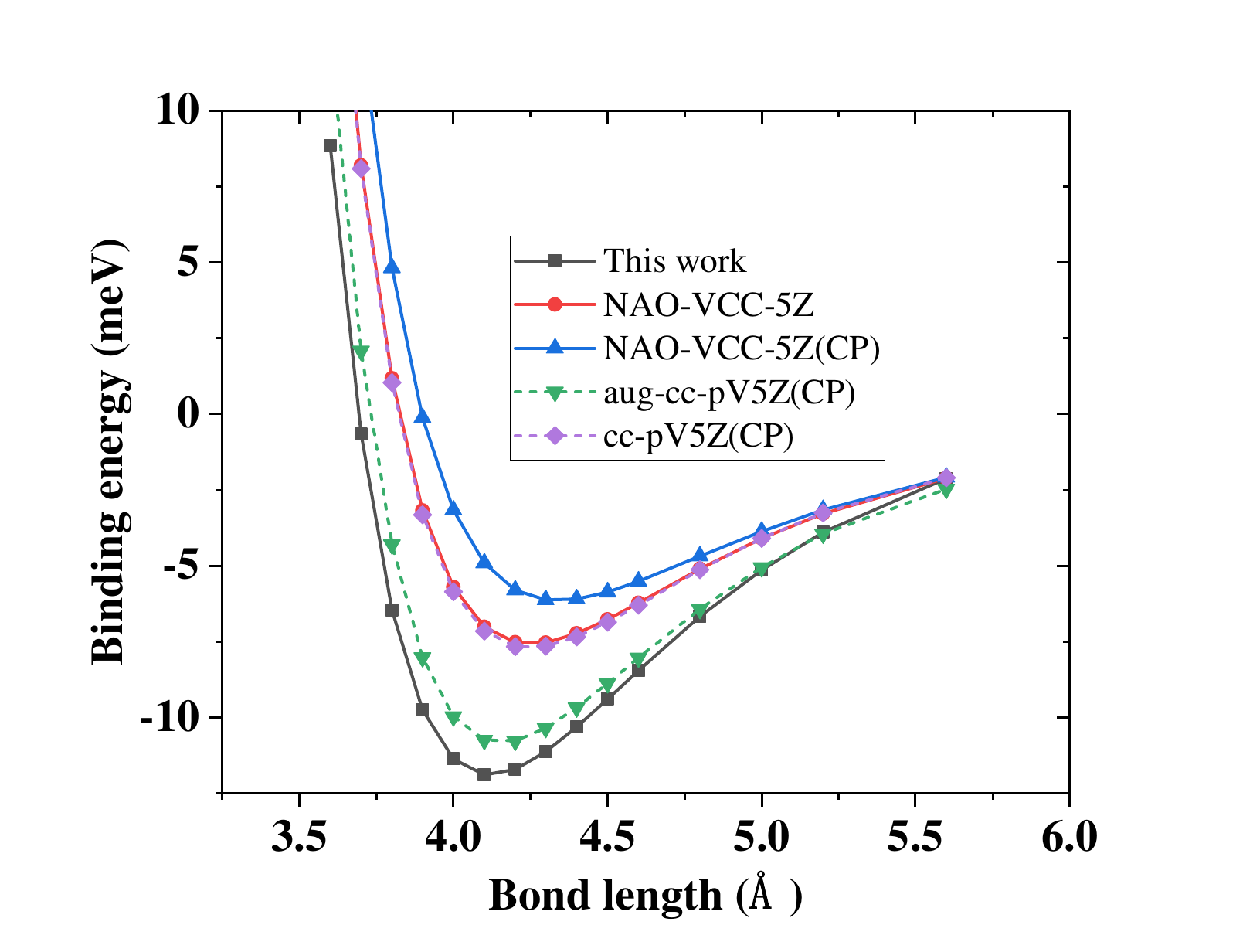}
    \caption{\label{fig:Kr2}  RPA@PBE binding energy curves of Kr$_2$ obtained using different basis sets, in comparison with the reference curves 
    obtained using the approach developed in this work. BSSEs are corrected for cc-pV5Z and aug-cc-pV5Z basis sets; 
    for NAO-VCC-5Z, both CP-corrected and uncorrected results are presented. The FC approximation is used for all calculations. }
    
\end{figure}

As alluded to above, our approach provides unambiguous reference results when finite basis set results show large
scatters. This applies to the case of Kr$_2$ dimer bound by pure van der Waals interactions. In Fig.~\ref{fig:Kr2},
we present the binding energy curves obtained using newly developed numerical NAO-VCC-$n$Z basis sets \cite{Yang/Zhang/Ren:2024},
as well as Dunning's cc-pV$n$Z and aug-cc-pV$n$Z \cite{dunning1989gaussian} basis sets. In the calculations presented here, the largest basis set available in each series
is used. The NAO-VCC-$n$Z was developed for periodic calculations, for which a small BSSE is a key requirement. However, this basis set seems to
significantly underbind Kr$_2$, regardless of whether the CP correction is performed. The GTO aug-cc-pV5Z basis set with
CP correction provides much improved results over NAO-VCC-5Z or cc-pV5Z, indicating the importance of including diffuse functions. However,
both aug-cc-pV5Z and cc-pV5Z suffer from significant BSSEs, and as such, the BSSE-uncorrected results produce too strong binding and fall off
the scale in Fig.~\ref{fig:Kr2}. Thus, a reasonable estimate of the CBS limit cannot be obtained by
comparing the CP-corrected and uncorrected results \cite{Yang/Zhang/Ren:2024}. Our reference results eventually establish that the largest aug-cc-pV5Z basis with CP corrections provides a rather good estimate of the RPA binding energies of Kr$_2$. 

Finally, we present in Table~\ref{tab:mol_dimer} the accurate binding energies of a set of diatomic molecules as predicted by the Sternheimer approach developed in this work. The set of molecules are chosen to be representative, composed of elements from light to heavy and bound together via
different interaction types.
For light elements up to Ar, AE calculations are performed, while for elements starting with
K, inner-core {\XB{electrons}} are frozen. We showed above that binding energies obtained from such
FC calculations are essentially identical to AE calculations.
For comparison, we also include the accurate results reported by Humer \textit{et al.} 
in Ref.~\citenum{humer2022approaching}, obtained using the explicitly correlated dRPA-F12 method \cite{Hehn/Tew/Klopper:2015} with the GTO basis set. Unfortunately,
such accurate results are only available for a few molecules containing light elements. 
Table~\ref{tab:mol_dimer} shows that, with an MAE of only 0.05 kcal/mol, dRPA-F12 method and the Sternheimer approach agree with each other remarkably well, confirming the excellent performance of the former in mitigating the one-electron BSIE. Note that in this work, our focus is on the basis error of RPA. Regarding the performance of RPA itself compared to more sophisticated methods and/or experiments, the readers are referred to Refs.~\cite{Ren/etal:2013,Pham/etal:2024} and references therein for further information.

\begin{table}[!h]
\caption{RPA@PBE binding energy (in kcal/mol) for a set of diatomic molecules.  The third column represents the results obtained using the Sternheimer approach developed in the present work. The fourth column lists very accurate results taken from the work of Humer \textit{et al} \cite{humer2022approaching}, 
obtained using explicitly correlated dRPA-F12 method. 
The mean absolute error (MAE) is measured for molecules where the dRPA-F12 results are available.}
\begin{tabular}{c c c c c }
\hline
\textbf{Molecule} & \textbf{Bond length  } & \textbf{This work} & \textbf{Ref.~\cite{humer2022approaching}} \\ \hline
H$_2$ & 0.74 & 108.72  & 108.69(-0.03) \\
He$_2$ & 4.30 &  ~0.001  &  / \\
Li$_2$& 2.70 & ~18.84   &~18.91(0.07)   \\
N$_2$ & 1.10 &  ~224.40 & 224.48(0.08) \\
F$_2$ & 1.43 & ~30.61   & ~30.61(0.00)  \\
LiH & 1.60 &   ~54.66    &  54.48(-0.16)    \\
HF &  0.92     &  132.77  &   132.78(0.01) \\
LiF & 1.58  & 127.36  & 127.36(0.00)\\
CO  &1.14  &  245.61  &245.65(0.04) \\
O$_2$  &  1.21  &113.79 & 113.74(-0.05)\\
Ne$_2$& 3.24 & ~~0.041 & / \\
Na$_2$ &  3.18  &~14.23 & /\\    
P$_2$ &1.91 &117.19 & / \\
Cl$_2$ &2.02 & ~50.18&/ \\
NaCl & 2.40  &  ~90.24&/  \\
Ar$_2$ & 3.84 & ~0.204 & / \\
K$_2$ &4.14 & ~10.32&/\\
Cu$_2$ &2.21 &~39.78 &/ \\
ZnO &1.72 & ~34.45&/ \\
Kr$_2$ & 4.10 & ~0.303 & / \\
As$_2$ &2.06 &~91.10&/ \\
Rb$_2$  &4.50 &~~9.18&/\\
Ag$_2$ &2.53& ~34.43 & /\\
Au$_2$ &2.53 & ~~64.11 & /\\
      \hline
      \hline
      MAE    &  /  & /&  0.05\\  
      \hline
      \hline
     \label{tab:mol_dimer}
\end{tabular}
\end{table}


In Table~S4 of the SM, we further present the FC-RPA binding energies for a larger set of
diatomic molecules, obtained using the Sternheimer approach and conventional schemes
with the largest available GTO cc-pV6Z/cc-pV5Z basis set, one still has an MAE of 1.60 kcal/mol.   Fortunately, by extrapolating to the CBS limit, the BSIE can be significantly reduced to below 0.4 kcal/mol, for both GTO and NAO basis sets.However, for heavy elements, such correlation-consistent NAO basis sets are unavailable, while the performance of currently available GTO basis sets developed for heavy elements for RPA calculations remains largely unexplored. In contrast, our basis-error-free RPA results provide 
invaluable references for developing high-quality AO basis sets for heavy elements. This possibility will be explored in future work.

We note that the numerical technique developed in the present work for correlated calculations without basis errors not only applies to RPA, but is also
directly extendable to any method based on density response functions $\chi^0$. One prominent example is the SOS-MP2 approach \cite{grimme2003improved,jung2004scaled}, which can be calculated as 
\begin{equation}
    E_\text{SOS-MP2} = - C_\text{OS}
     \int_{0}^{\infty} \frac{\mathrm{d} \omega}{2 \pi} \operatorname{Tr}\left[v \chi_{0}^{\uparrow} v \chi_{0}^{\downarrow}\right]
\end{equation}
where $\chi_{0}^{\sigma}$ with $\sigma = \uparrow, \downarrow$ is spin-resolved density response function and $C_{OS}=1.3$. 
The SOS-MP2 has been found to produce better
results compared to the standard MP2 in certain situations \cite{jung2004scaled} without the need of calculating the second-order exchange component.
In Sec.~VII of the SM, we show that numerically highly-precise SOS-MP2 correlation energy can be obtained using our technique, similar to the
RPA case.

\textit{Summary.}-In this work we developed a numerical approach that allows to perform basis-error-free calculations \textit{ab initio} correlated  methods that are formulated in terms of density response function. 
The present implementation is restricted to the RPA and SOS-MP2 methods and diatomic molecules, but extending this approach to more general situations is foreseeable. In Sec.~VIII of the SM, we showed how to extend this approach to polyatomic systems by solving the Sternhemier equation on three-dimensional self-adapted finite-element grids. The approach yields numerically precise absolute all-electron
correlation energies and is applicable to both light and heavy elements. It thus offers an unambiguous way to quantify 
the numerical uncertainty associated with the commonly used computational protocol in quantum chemistry, i.e., finite 
GTO basis set with extrapolation to the CBS limit, frozen-core approximations, and the counterpoise correction to
the BSSE. The reference results can also be used to guide the development of more efficient correlation-consistent AO basis sets, particularly for cases where such basis sets are not yet available.

\textit{Acknowledgment.}-We acknowledge the funding support by the Strategic Priority
Research Program of Chinese Academy of Sciences
under Grant No. XDB0500201 and by the National Key
Research and Development Program of China (Grant Nos.
2022YFA1403800 and 2023YFA1507004). This work was also supported by the National Natural Science Foundation of China (Grants Nos. 12134012, 12374067, and 12188101). 

\textit{End Matter on the Methodology.}—The key to basis-error-free RPA calculations is to obtain numerically accurate first-order KS wavefunction (WF) induced by arbitrary external perturbation. To introduce
our approach, we start with the KS Hamilitonian $H^{(0)}$,
\begin{equation}
    H^{(0)} = -\frac{1}{2} \nabla^2 + v_{\text{eff}}(\bm{r})
\end{equation}
where $v_{\text{eff}}(\bm{r})$ is the KS  effective potential. 
 Upon adding a small frequency-dependent perturbation $V^{(1)}(\bm{r})e^{i\omega t}$ to the Hamiltonian $H^{(0)}$, the linear response of the system is governed by the following frequency-dependent Sternheimer equation \cite{peng2023basis},
 \begin{equation}
			      (H^{(0)}-\epsilon_i+i\omega)\psi_{i}^{(1)}(\bm{r},i\omega)=(\epsilon_i^{(1)}-V^{(1)})\psi_i(\bm{r})
			      \label{eq:st}
		      \end{equation}
where $\psi_i(\bm{r})$ and $\epsilon_i$ are the KS orbitals and orbital energies, and  
$\psi_{i}^{(1)}(\bm{r},i\omega)$ and $\epsilon_i^{(1)}$ are their first-order variations, respectively. 
This 3D differential equation simplifies for diatomic molecules which have the rotational symmetry with respect to the axis connecting the two nuclei. 
To exploit this symmetry, we chose the prolate spheroidal coordinates system, whereby the two atoms are placed at the two foci of the spheroid, and
the position of a point in space is described by a set of variables $(\mu,\nu, \theta )$. These prolate spheroidal  variables are related to the Cartesian variables as
$x=\frac{R}{2}\text{sinh}(\mu)\text{sin}(\nu)\text{cos}(\theta)$, $y=\frac{R}{2}\text{sinh}(\mu)\text{sin}(\nu)\text{sin}(\theta)$, and $z=\frac{R}{2}\text{cosh}(\mu)\text{cos}(\nu)$,
with $R$ being the distance between the two atoms, and  $ 0\le \mu \le \infty$, $0\le  \nu  \le \pi$, $0\le  \theta \le 2\pi$. A graphical illustration of the geometrical
meaning of these variables and a discussion of the representation of the WFs and operators
in this coordinate system are provided in the Supplemental Material (SM) (see Fig.~S1 and
Sec.~I~A). 
In the prolate spheroidal coordinates system, the Hamiltonian and WFs adopt the following forms,
\begin{equation}
\label{eq:H^0}
    H^{(0)}=-\frac{1}{2}\nabla^2 +v_{\text{eff}}(\mu,\nu)
\end{equation}
\begin{equation}
\label{eq:0-order-wave}
\psi_i(\bm{r})=f_i(\mu,\nu)e^{im\theta}
\end{equation}
The simplification comes from the fact that the dependence of the physical quantities on the angular variable $\theta$ can be eliminated. 

A commonly used approach to evaluate the RPA correlation energy is the resolution of identity (RI) approximation \cite{Eshuis/Yarkony/Furche:2010,ren2012resolution}, under which 
the KS response function is represented as a matrix
within a set of auxiliary basis function (ABF) $\{P(\bm{r})\}$.  By taking the ABFs $\{P(\bm{r})\}$ as the external perturbations, one can formulate a RI-RPA computational scheme that is free of single-particle BSIE,
as discussed in Ref.~\cite{peng2023basis}. The idea follows from a similar BSIE correction scheme for $GW$ developed in the linearized augmented plane wave
framework \cite{betzinger2012precise,betzinger2013precise,betzinger2015precise}. The mathematical formulation of Sternheimer-based RI-RPA scheme for diatomic molecules is discussed in Sec.~I~B of the SM.

As shown in Ref.~\cite{peng2023basis}, the error due to the incompleteness of the ABFs is much smaller than that of the single-particle basis. In particular, the standard ABFs constructed on the fly in the FHI-aims code
\cite{ren2012resolution,ihrig2015accurate} is sufficiently good for most practical RPA calculations. Nevertheless, this error is still visible and it is desirable to eliminate the BSIE of the ABFs as well to achieve a numerical precision of the absolute RPA correlation energy to the meV level.
To this end, we use the iterative diagonalization method to directly determine the eigenspectra of the operator $\chi^0(i\omega)v$. This approach has been used to calculate the dielectric function in the pseudopotential plane-wave context \cite{wilson2008efficient,wilson2009iterative,nguyen2009efficient}.  The key difference
here is that we solve the Sternheimer equation on a non-uniform grid and fully exploit the
symmetry of diatomic systems. Only by doing so are converged all-electron RPA calculations possible for molecules.

Specifically, we start with a trial eigenfunction $\phi(\bm{r})$ which 
is set in the following form to account for the prolate spheroidal  symmetry,
\begin{equation}
    \phi(\bm{r})=\phi(\mu,\nu)e^{iM\theta}\, .
    \label{eq:trial_func}
\end{equation}
Now we need to determine the first-order density $\Delta n(\bm{r})$ upon applying $\chi^0(i\omega)v$ to the trial function $\phi(\mu,\nu)e^{iM\theta}$, i.e.,
\begin{equation}
    \chi^0(i\omega)v \phi(\mu,\nu)e^{iM\theta}=\Delta n(\bm{r},i\omega)=\Delta n(\mu,\nu,i\omega)e^{iM\theta}\, ,
    \label{eq:chi0vphi}
\end{equation}
and then take the resultant $\Delta n(\bm{r},i\omega)$ as the input vector of the next iteration.  The process is repeated until $\Delta n(\bm{r},i\omega)$  is
converged within a given threshold. By doing so a pair of eigenvalue and eigenvector of the $\chi^0(i\omega)v$ operator is obtained. 

We further note that the action of the combined $\chi^0(i\omega)v$ operator on a function of the form given in Eq.~\ref{eq:trial_func} can be executed in two successive steps. 
First, applying the Coulomb operator $v$ on $\phi(\bm{r})$ amounts to computing the Hartree potential corresponding to a density distribution of $\phi(\bm{r})$,
\begin{equation}
 \phi_h(\bf{r})= \int \frac{\phi(\bf{r^\prime})}{|\bf{r} - \bf{r^\prime}|} d\bf{r^\prime} \, .
\label{eq:hartree}
\end{equation}
In practice, $\phi_h(\bf{r})$ is computed by solving the Poisson equation on the prolate spheroidal coordinates system. 
In the second step, consider the resultant $\phi_{h}(\mu,\nu)$ as a perturbation and solve the Sternheimer equation to determine the first-order WF $f_{i}^{(1)}(\mu,\nu,i\omega)$. Then, the variation of the electron density is given by
\begin{equation}
    \Delta n(\mu,\nu,i\omega)=\sum_{i}^{\text{occ}}f_i^*(\mu,\nu)*f_i^{(1)}(\mu,\nu,i\omega) + c.c.\, ,
    \label{eq:1-order-density-radial}
\end{equation}
where $c.c.$ denotes the complex conjugate of the former term. Following these two steps, the result by applying the $\chi^0(i\omega)v$ operator on an input vector can be obtained. $\Delta n(\mu,\nu,i\omega)$ provides the input vector
for the next iteration. This process is repeated until the output and input vectors are parallel to each other. 
More rigorous derivations and implementation details are given in Sec.~I~C of the SM. 

In this work, we use the package ARPACK \cite{lehoucq1998arpack} based on the Arnoldi algorithm \cite{saad1981krylov}, together with the above-described two-step procedure,  to iteratively diagonalize $\chi^0(i\omega)v$. 
In particular, ARPACK can provide a specified number of eigenvalues arranged in the desired ascending or descending order. In the present case, $\chi^0(i\omega)v$ is a negative definite operator, and all eigenvalues are negative. 
Once the eigenspectrum of the $\chi^0(i\omega)v$ is determined, the the RPA correlation energy for a diatomic molecule 
can be calculated using Eq.~\ref{eq:EcRPA}.
In practical calculations, we set the parameter for ARPACK so that it will output the eigenvalues in order of their magnitude, i.e. their contributions to the RPA correlation energy. As discussed previously based on Eq.~\ref{eq:EcRPA}, by systematically increasing $N_{\text{eigen}}$ and $M$, the RPA correlation energy can be obtained to any desired precision for diatomic molecules.

Our approach is implemented in the all-electron FHI-aims code package \cite{blum2009ab}. The employment of numerical atomic orbital (NAO) basis set enables an accurate description of the manifold of occupied states, including the
core one. This is also an essential factor that makes the high-precision all-electron RPA calculations possible. Further implementation details are given in Sec.~II of the SM.
For heavy elements with $Z>20$, the relativistic effect is treated under the atomic zeroth-order regular approximation (ZORA) \cite{Lenthe/Baerends/Snijders:1994}, as implemented in FHI-aims \cite{blum2009ab}.
\renewcommand{\thefigure}{S\arabic{figure}}
\setcounter{figure}{0}

\renewcommand{\thetable}{S\arabic{table}}
\setcounter{table}{0}

\renewcommand{\theequation}{S\arabic{equation}}
\setcounter{equation}{0}





\clearpage

\ifdefined\onecolumngrid
  \onecolumngrid
\fi

\clearpage

\begin{widetext}
\begin{center}
   \textbf{Supplemental materials for}\\[6pt]
  { \textbf{Ab initio correlated calculations without basis set error:\\
    Numerically precise all-electron RPA correlation energies for diatomic molecules}}\\[12pt]

  { Hao Peng\textsuperscript{1,2} and Xinguo Ren\textsuperscript{1,*}}\\[6pt]

  {\small
    \textsuperscript{1}Institute of Physics, Chinese Academy of Sciences, Beijing 100190, China\\
    \textsuperscript{2}University of Chinese Academy of Sciences, Beijing 100049, China
  }\\[6pt]
\end{center}
\end{widetext}


\section{Numerical methods}
\subsection{Prolate spheroidal coordinates system}
\begin{figure}[htbp]
    
    \centering
    \includegraphics[scale=0.35]{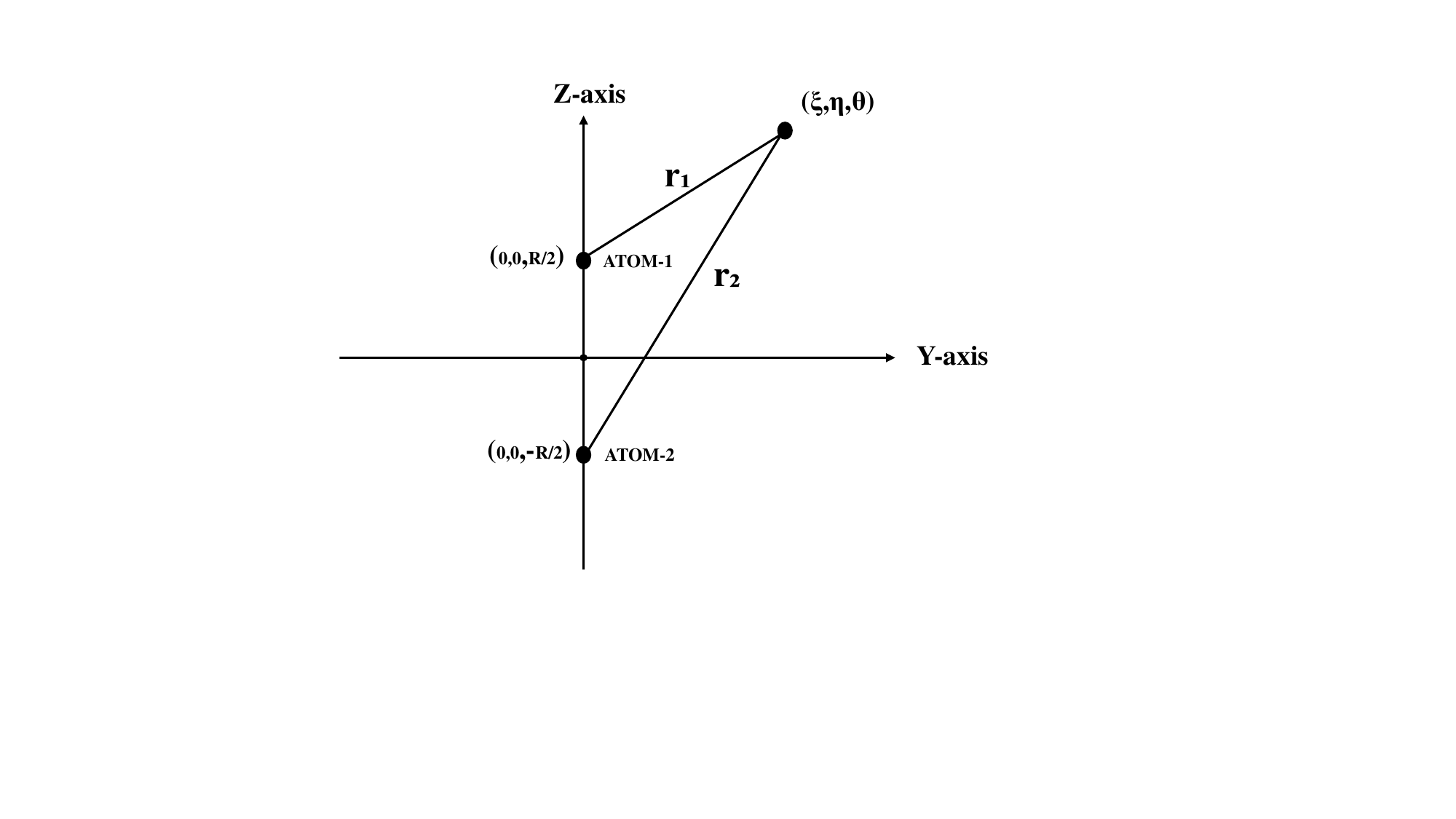}
    \caption{\label{fig:coord}Schematic diagram of prolate spheroidal coordinates system}
    
\end{figure}

Diatomic molecule systems can be described in prolate spheroidal coordinates \cite{kobus2013finite,aubert1974prolate,kereselidze2015coulomb,becke1982numerical}.
\begin{equation}
\begin{aligned}
&\xi=\frac{(r_1+r_2)}{R} \quad \quad 1\le \xi \le \infty \\
&\eta=\frac{(r_1-r_2)}{R} \quad \ -1\le \eta \le 1 \\
&\theta \quad \quad \quad \quad \ \quad \quad \quad 0\le \theta \le 2\pi
\end{aligned}
\end{equation}
where $r_1$ and $r_2$ represents the distance from a given point to two
atomic centers, $R$ represents the distance between the two atomic centers. Furthermore,
$\xi$  describes the distance from a given point to the entire diatomic 
system, $\eta$ characterizes the angle between the line connecting a given point and the whole diatomic system and the line connecting the diatomic system. $\theta$ describes the rotation angle along the axis of the line connecting two atomic centers. Figure.~\ref{fig:coord} illustrates 
how to characterize a spatial point in the prolate spheroidal coordinates system.

In order to describe more accurately the wave function and potential in the vicinity of atomic nuclei, the prolate spheroidal coordinates  $(\xi, \eta , \theta )$
are transformed into $(\mu,\nu, \theta )$ variables,
\begin{equation}
\begin{aligned}
&\mu = \operatorname{arccosh} \, \xi, \quad 0 \leq \mu \leq \infty, \\
&\nu = \arccos \, \eta, \quad\ \ 0 \leq \nu \leq \pi.
\end{aligned}
\label{eq:ellipsoidal_uniform}
\end{equation}
whereby a uniform grid of $(\mu,\nu, \theta )$ translates into a gradually denser 
grids of $(\xi, \eta , \theta )$ in the vicinity of the atomic nuclei.
Considering the rotational symmetry of diatoms with respect to the axis of the connecting line, 
the Kohn-Sham (KS) wave function of a diatomic molecule can separate variables as,
\begin{equation}
\psi(\mu,\nu,\theta)=f(\mu,\nu)e^{im\theta}\, .
\end{equation}
The ground-state density of the system can be represented as,
\begin{equation}
\begin{aligned}
    n(\mu,\nu,\theta)&=\sum_{i}^{\text{occ}}\psi_i^*(\mu,\nu,\theta)\psi_i(\mu,\nu,\theta)\\
    &=\sum_{i}^{\text{occ}}f_i^*(\mu,\nu)*f_i(\mu,\nu)\\ 
    &=n(\mu,\nu)
\end{aligned}
	\end{equation}
Thus the density is independent of $\theta$, and so does the effective potential $v_{\text{eff}}$,
\begin{equation}
		v_{\text{eff}}(\mu,\nu,\theta)=v_{\text{eff}}(\mu,\nu)\, .
	\end{equation}
 In the prolate spheroidal coordinates , the Laplacian operator acting on a wavefunction (WF) can be expressed as\cite{kobus2013finite},
 \begin{equation}
 \label{eq:Nabula}
 \begin{aligned}
     \nabla^2 f(\mu,\nu)e^{im\theta} =&\frac{4}{R^2(\xi^2-\eta^2)}[
     \frac{\partial^2 f(\mu,\nu)}{\partial \mu^2}+\\ &\frac{\xi}{\sqrt{\xi^2-1}}\frac{\partial f(\mu,\nu)}{\partial \mu }+\frac{\partial^2 f(\mu,\nu)}{\partial \nu^2}+\\ &\frac{\eta}{\sqrt{1-\eta^2}}\frac{\partial f(\mu,\nu)}{\partial \nu}-\\
     &m^2(\frac{1}{\xi^2-1}+\frac{1}{1-\eta^2})f(\mu,\nu)] e^{im\theta} \, .
\end{aligned}
\end{equation}
 Because of the rotational symmetry, $\theta$ is a trivial coordinate, and hence in Eq.~\ref{eq:Nabula}, only the derivatives of $\mu$ and $\nu$ appear.
Define $\nabla_{\mu,\nu}^2$,
\begin{equation}
 \label{eq:Nabula_ellip}
 \begin{aligned}
     \nabla_{\mu,\nu}^2 =&\frac{4}{R^2(\xi^2-\eta^2)}(\frac{\partial^2 }{\partial \mu^2}+\frac{\xi}{\sqrt{\xi^2-1}}\frac{\partial }{\partial \mu }+\\ 
     &\frac{\partial^2 }{\partial \nu^2}+\frac{\eta}{\sqrt{1-\eta^2}}\frac{\partial }{\partial \nu}\\
     &-m^2(\frac{1}{\xi^2-1}+\frac{1}{1-\eta^2}) \, ,
\end{aligned}
\end{equation}
one can easily arrive at,
\begin{equation}
 \label{eq:Nabula_2}
     \nabla^2 f(\mu,\nu)e^{im\theta} =e^{im\theta} . \nabla_{\mu,\nu}^2f(\mu,\nu)
\end{equation}
As can be seen, $\theta$ becomes a trial variable in the prolate spheroidal coordinates  system and can be eliminated from actual calculations. This allows us to describe the spatial distribution of physical quantities of diatomic molecules using only two variables $\mu,\nu$. We set dense grids in both variables to form a complete Hilbert space. This makes our calculations free of
basis set incompleteness error (BSIE), achieving unprecedented accuracy.

 \subsection{Sternheimer equation in prolate spheroidal coordinates  system}
 We start with the single-particle KS Hamiltonian $H^{(0)}$ ,which in general has the following form,
\begin{equation}
    H^{(0)}=-\frac{1}{2}\nabla^2 +v_{\text{eff}}(\bm{r})
\end{equation}
where $v_{\text{eff}}(\bm{r})$ is the KS  effective potential. 
 After adding a small frequency-dependent perturbation $V^{(1)}(\bm{r})e^{i\omega t}$ to the Hamiltonian $H^{(0)}$, the linear response of the system is governed by the following frequency-dependent Sternheimer equation\cite{marques2012fundamentals,sternheimer1954electronic,sternheimer1957electronic,sternheimer1970quadrupole,mahan1980modified,zangwill1980density,mahan1982van},
 \begin{equation}
			      (H^{(0)}-\epsilon_i+i\omega)\psi_{i}^{(1)}(\bm{r},i\omega)=(\epsilon_i^{(1)}-V^{(1)})\psi_i(\bm{r})
			      \label{eq:st}
		      \end{equation}
where $\psi_i(\bm{r})$ and $\epsilon_i$ are the KS orbitals and orbital energies, and  
$\psi_{i}^{(1)}(\bm{r},i\omega)$ and $\epsilon_i^{(1)}$ are their first-order variations, respectively.
In the prolate spheroidal coordinates  system, the Hamiltonian and WFs adopt the following forms,
\begin{equation}
\label{eq:H^0}
    H^{(0)}=-\frac{1}{2}\nabla^2 +v_{\text{eff}}(\mu,\nu)
\end{equation}
\begin{equation}
\label{eq:0-order-wave}
\psi_i(\bm{r})=f(\mu,\nu)e^{im\theta}
\end{equation}
The simplification comes from the fact that the dependence of the physical quantities on the angular variable $\theta$ can be eliminated. 

Within the resolution of identity (RI) formulation of the random phase approximation (RPA) \cite{ren2012resolution}, one can
take the auxiliary basis functions (ABFs) as the external perturbation. 
This approach enables the attainment of the RI-RPA correlation energies free of single-particle basis set incompleteness error (BSIE) \cite{peng2023basis}. 
In terms of prolate spheroidal coordinates ,  an ABF $P(\bm{r})$, given by a radial function multiplied by spherical harmonics, can be expressed as,
	\begin{equation}
		 P(\bm{r})=V^{(1)}(\bm{r})=V^{(1)}(r)Y_L^M(\theta,\phi)=V^{(1)}(\mu,\nu)\, e^{iM\theta} \, . 
		\label{eq:auxil_P}
	\end{equation}
Eqs.~\ref{eq:0-order-wave} and \ref{eq:auxil_P} show that the zeroth-order WF and the external perturbative potential have the same structure, where the angular dependence
can be separated out as a simple phase factor. Utilizing
this property, we can assume, without losing generality, the following form of the first-order WF,  
\begin{equation}
\label{eq:1-order-wave}
\psi_i^{(1)}(\bm{r},i\omega)=\sum_{m^\prime}f_{i,m^\prime}^{(1)}(\mu,\nu,i\omega)e^{im^\prime \theta}\, .
\end{equation}
In the above equations, we use $m$, $m^\prime$ and $M$ to label the angular quantum number of the zeroth-order and first-order WFs, and the external perturbation (ABF), respectively.

Combining Eqs.~\ref{eq:st},  \ref{eq:0-order-wave}, \ref{eq:auxil_P}, and \ref{eq:1-order-wave}, we can obtain, by separating variables, the two-dimensional (2D) radial Sternheimer equation independent of $\theta$,
\begin{equation}
\label{eq:radial st}
\begin{aligned}
    &(-\frac{1}{2}\nabla_{\mu,\nu}^2 +v_{\text{eff}}(\mu,\nu)-\epsilon_i+i\omega)f_{i,m^\prime}^{\,(1)}(\mu,\nu,i\omega) \\
    &=  \delta_{m+M,m^\prime}(\epsilon_i^{(1)}\delta_{m,m^\prime}-V^{(1)}(\mu,\nu))\,f_i(\mu,\nu) \, .
\end{aligned}
	\end{equation}
The derivation of Eq.~\ref{eq:radial st} is given below. 
Using Eqs.~\ref{eq:H^0}-\ref{eq:0-order-wave}, \ref{eq:auxil_P} and \ref{eq:1-order-wave}, the $l.h.s.$ and $r.h.s.$ of Eq.~\ref{eq:st} become,
\begin{equation}
    l.h.s.=(-\frac{1}{2}\nabla^2+v_{\text{eff}}(\mu,\nu)-\epsilon_i+i\omega)
\left[\sum_{ m^\prime}f_{i,m^\prime}^{(1)}(\mu,\nu,i\omega)e^{im^\prime \theta}\right] 
\label{eq:radial st left1} \\
\end{equation}
\begin{equation}
    r.h.s.=[\epsilon_i^{(1)}-V^{(1)}(\mu,\nu)e^{iM\theta}]f_i(\mu,\nu)e^{im\theta}\ ,
\label{eq:radial st right}
\end{equation}
respectively.

According to Eq.~\ref{eq:Nabula_2}, Eq.~\ref{eq:radial st left1} further changes to,
\begin{equation}
\begin{aligned}
    & l.h.s.= \\ 
    & \sum_{m^\prime}e^{im^\prime \theta}  (-\frac{1}{2}\nabla_{\mu,\nu}^2+v_{\text{eff}}(\mu,\nu)-\epsilon_i+i\omega) f_{i,m^\prime}^{(1)}(\mu,\nu,i\omega)
\end{aligned}
\end{equation}
Multiplying both sides of the equation with $e^{im^{\prime\prime}\theta}$ and integrating with respect to the angular coordinates ${\theta}$, and further utilizing the orthogonality relationship,
\begin{equation}
\int{\rm d}{\theta} e^{im\theta}e^{im^\prime \theta}=2\pi \delta_{mm^{\prime}}\
\end{equation}
one obtains,
\begin{equation}
\begin{aligned}
   & l.h.s.= \\
   & 2\pi\sum_{m^\prime}\delta_{m^{\prime \prime}m^{\prime}}  (-\frac{1}{2}\nabla_{\mu,\nu}^2+v_{\text{eff}}(\mu,\nu)-\epsilon_i+i\omega) f_{i,m^\prime}^{(1)}(\mu,\nu,i\omega)\, ,
\end{aligned}
\label{eq:radial_lhs} 
\end{equation}
and
\begin{equation}
    r.h.s.= 2\pi[\epsilon_i^{(1)}\delta_{mm^{\prime \prime}} -V^{(1)}(\mu,\nu)\delta_{m+M,m^{\prime \prime}}]f_i(\mu,\nu)\, .
\label{eq:radial st right 2}
\end{equation}

In the prolate spheroidal coordinates  system, the computation of $\epsilon^{(1)}$ is reduced from a three-dimensional (3D) integral to an integral over $\mu,\nu$ times an integral over $\theta$,
\begin{equation}
\begin{aligned}
  & \epsilon_i^{(1)}=\left \langle \psi_i\right |V^{(1)} \left | \psi_i \right \rangle \\
  &=\iint {\rm d} {\mu} {\rm d} {\nu} f_i^*(\mu,\nu)V^{(1)}(\mu,\nu)f_i(\mu,\nu)\int {\rm d} {\theta} e^{-im\theta}e^{iM\theta}e^{im\theta}\\
  &=\epsilon_i^{(1)}\delta_{M0} \, .
\end{aligned}
\end{equation}
That is, only the first-order energy corresponding to a perturbation of $M=0$ is nozero.
Therefore, utilizing $\delta_{M0}\delta_{mm^{\prime\prime}}=\delta_{M0}\delta_{m+M,m^{\prime\prime}} $, Eq.~\ref{eq:radial st right 2} changes to
\begin{equation}
\label{eq:radial_rhs}
    r.h.s.= 2\pi \delta_{m+M,m^{\prime \prime}}[\epsilon_i^{(1)}\delta_{mm^{\prime \prime}} -V^{(1)}(\mu,\nu)]f_i(\mu,\nu)\, .
\end{equation}
Finally, by equating Eq.~\ref{eq:radial_lhs} and \ref{eq:radial_rhs} and replacing $m^{\prime \prime} $ 
by $m^{\prime} $, one obtains the desired 2D Sternheimer equation given by Eq.~\ref{eq:radial st},
which suggests that the equation has nontrivial solutions if and only if $m^\prime = m+M$. 
Hence, Eq.~\ref{eq:1-order-wave} can be simplified as
\begin{equation}
\label{eq:1-order-wave_simp}
\psi_i^{(1)}(\bm{r},i\omega)=f_{i,m+M}^{(1)}(\mu,\nu,i\omega)e^{i(m+M)) \theta}\, .
\end{equation}
Solving Eq.~\ref{eq:radial st} on a dense prolate spheroidal   grid, one attains the only surviving component $f_{i,m^\prime}^{\,(1)}(\mu,\nu,i\omega)$ in the
first-order WF $\psi_i^{(1)}(\bm{r},i\omega)$ without single-particle BSIE. 
With the numerically precise first-order WF, we then proceed to calculate the accurate density response function represented in terms of the auxiliary basis and the RI-RPA correlation energy that is also free of single-particle BSIE.
In this way, similar to what is done in Ref.~\cite{peng2023basis} for atoms, we can now calculate highly accurate RI-RPA energies for diatomic molecules.

\subsection {Iterative diagonalization of density response function}
As shown in Ref.~\cite{peng2023basis}, the error due to the incompleteness of the ABFs is much smaller than that of the single-particle basis. In particular, the standard ABFs constructed on the fly in the FHI-aims code
\cite{ren2012resolution,ihrig2015accurate} is sufficiently good for most practical RPA calculations. Nevertheless, this error is still visible and, to achieve a numerical precision of the absolute RPA correlation energy up to the meV level, it is necessary to eliminate the BSIE of the ABFs as well.
To this end, we use the iterative diagonalization method to directly determine the eigenspectra of the operator $\chi^0(i\omega)v$.
Specifically, we start with a trial eigenfunction $\phi(\bm{r})$, which, considering the prolate spheroidal symmetry,  can be set in the following form,
\begin{equation}
    \phi(\bm{r})=\phi(\mu,\nu)e^{iM\theta}\, .
    \label{eq:trial_func}
\end{equation}
The essential step of iterative diagonalization is to repeatedly apply the operator to a vector until
the resultant vector and the input vector are parallel to each other. In the present case, we need to determine the first-order density $\Delta n(\bm{r})$ upon applying $\chi^0(i\omega)v$ to the trial function $\phi(\mu,\nu)e^{iM\theta}$, i.e.
\begin{equation}
    \chi^0(i\omega)v \phi(\mu,\nu)e^{iM\theta}=\Delta n(\bm{r},i\omega)=\Delta n(\mu,\nu,i\omega)e^{iM\theta}\, ,
\end{equation}
and then take the resultant $\Delta n(\mathbf{r},i\omega)$ as the input vector for the next iteration.  The process is repeated until $\Delta n(\bm{r},i\omega)$  is
converged within a given threshold. By doing so, a pair of eigenvalue and eigenvector of the $\chi^0(i\omega)v$ operator is obtained.
A nice feature to note is that the distribution of first-order density in the $\theta$ direction is consistent with the perturbation (trial function),
\begin{equation}
    \begin{aligned}
        \Delta n(\mathbf{r},i\omega) &= \sum_i^{\text{occ}} \psi_i^*(\mathbf{r})*\psi_i^{(1)}(\mathbf{r},i\omega) +\sum_i^{\text{occ}} \psi_i(\mathbf{r})*{\psi_i^*}^{(1)}(\mathbf{r},i\omega)\\
        &=\sum_i^{\text{occ}}f_i^*(\mu,\nu)e^{-im\theta}*f_{i}^{(1)}(\mu,\nu,i\omega)e^{i(m+M)\theta}\\
        &+\sum_i^{\text{occ}}f_i(\mu,\nu)e^{im\theta}*{f_{i}^*}^{(1)}(\mu,\nu,i\omega)e^{i(-m+M)\theta}\\
        &=\sum_i^{\text{occ}}f_i^*(\mu,\nu)*f_{i}^{(1)}(\mu,\nu,i\omega)e^{iM\theta}\\
        &+\sum_i^{\text{occ}}f_i(\mu,\nu)*{f_{i}^*}^{(1)}(\mu,\nu,i\omega)e^{iM\theta}\\
        &=\Delta n(\mu,\nu,i\omega)e^{iM\theta}\, .
    \end{aligned}
\end{equation}
Thus the iterative diagonalization can be done independently for different angular momentum channels.

We note that the action of the combined $\chi^0(i\omega)v$ operator on a function of the form given in Eq.~\ref{eq:trial_func} can be executed in two successive steps. 
First, applying the Coulomb operator $v$ on $\phi(\bm{r})$ amounts to computing the Hartree potential corresponding to a density distribution of $\phi(\bm{r})$,
\begin{equation}
 \phi_h(\bf{r})= \int \frac{\phi(\bf{r^\prime})}{|\bf{r} - \bf{r^\prime}|} d\bf{r^\prime} \, .
\label{eq:hartree}
\end{equation}
However, for the trial function of the form in Eq.~\ref{eq:trial_func}, the integration of Eq.~\ref{eq:hartree} cannot be reduced to one-dimensional integration like the atomic problem. To avoid complicated three-dimension integration, we choose to solve the Poisson equation on prolate spheroidal coordinates system to attain the Hartree potential,
\begin{equation}
    \nabla^2 \phi_{h}(\bf{r})=-4\pi \phi(\bf{r})\, .
\end{equation}
Since $e^{iM\theta}$ is an eigenfunction of the $\nabla^2$ operator, $\phi_h(\bf{r})$ has the same $\theta$-dependence as $\phi(\bf{r})$, i.e.
\begin{equation}
    \phi_h(\bm{r})=\phi_h(\mu,\nu)e^{iM\theta}\, .
\end{equation}

According to Eq.~\ref{eq:Nabula_2}, the Poisson equation becomes,
\begin{equation}
    \nabla_{\mu,\nu}^2 \phi_{h}(\mu,\nu)=-4\pi \phi(\mu,\nu)
    \label{eq:Pos_epli}
\end{equation}

In the second step, consider the obtained $\phi_{h}(\mu,\nu)$ as a perturbation to the system, and solve the Sternheimer equation to determine the first-order WF $f_{i,m^\prime}^{\,(1)}(\mu,\nu,i\omega)$. From Eq.~\ref{eq:1-order-wave} and noticing that only the component of
$m'=m+M$ is nonzero, we have
\begin{equation}
\label{eq:1-order-wave1}
\psi_i^{(1)}(\bm{r},i\omega)=f_{i,m+M}^{(1)}(\mu,\nu,i\omega)e^{i(m+M)\theta}\, .
\end{equation}
Then, denoting $f_{i}^{(1)}(\mu,\nu,i\omega)=f_{i,m+M}^{(1)}(\mu,\nu,i\omega)$, the
first-order change of the electron density in the prolate spheroidal coordinates is given by 
\begin{equation}
\begin{aligned}
    \Delta n(\mu,\nu,i\omega)&=\sum_{i}^{\text{occ}}f_i^*(\mu,\nu)*f_i^{(1)}(\mu,\nu,i\omega)\\
    &+ \sum_{i}^{\text{occ}}f_i(\mu,\nu)*{f_i^*}^{(1)}(\mu,\nu,i\omega).
    \label{eq:1-order-density-radial}
    \end{aligned}
\end{equation}
$\Delta n(\mu,\nu,i\omega)$ will be used as the perturbation to the system in the next iteration.

\section{Implementation details}
\subsection{Obtaining Hamiltonian on real space grids}
In our work, we first use FHI-aims \cite{blum2009ab} to perform all-electron DFT \cite{kohn1965self} calculations based on the PBE \cite{perdew1996generalized} functional to obtain the occupied  KS eigenvctors and KS eigenvalues of the system. FHI-aims yields the KS eigenvectors expanded in terms of numerical atomic orbital (NAO) basis set, but in our work, we need to provide the KS wavefunction and effective potentials in real space prolate spheroidal coordinates. Our approach is to interpolate the NAO basis functions and their values to the prolate spheroidal  grid points. Since the angular part of
the NAO is just the spherical harmonics, we only need to perform a one-dimensional radial interpolation. Considering the distance between each prolate spheroidal  grid point and the atom on the basis function is centering and using one-dimensional cubic spline interpolation, we can retrieve accurately the value of the basis functions on the prolate spheroidal  grid points.  Using the KS eigenvector to linearly combine the interpolated basis functions, all occupied  KS wave functions on the prolate spheroidal  grid can be evaluated. Then, by adding up the occupied states, the charge density on the prolate spheroidal  grids can be obtained. Next, based on the density, the effective potential on the grids can be reconstructed, following the principle that the effective potential is the functional of the density. This interpolation technique can help us obtain the Hamiltonian 
represented on any real space grids, not limited to the prolate spheroidal  grid here.

\subsection{Grids finite difference and  sparse matrix linear equation system}
\label{appendix_fd}
For the grids setting and difference and integration techniques in the prolate spheroidal coordinates system, we have referred to Ref~\cite{kobus2013finite}.  \\

 We set the  uniform grid points in the $\mu$,$\nu$  direction, 
\begin{equation}
 \begin{aligned}
      &\mu(i)=(i-\frac{1}{2})h_{\mu} \quad i=1,2,3...,N_{\mu}\\
     &\nu(i)=(i-\frac{1}{2})h_{\nu}\quad i=1,2,3...,N_{\nu}
 \end{aligned}
 \end{equation}
 which, via Eq.~\ref{eq:ellipsoidal_uniform}, automatically transforms to a set of non-uniform
 grid points which are dense in the near-kernel region while sparse in the far-kernel region. 
Here, $N_\mu$ and $N_\nu$ usually take around 200, and $h_\nu=\frac{\pi}{N_\nu}$, $h_\mu=\frac{\mu_{\infty}}{N_{\mu}}$. $\mu_{\infty}$ has the following relationship with the the infinite distance $r_{\infty}$ we  choose in actual calculation,
\begin{equation}
    \mu_{\infty} = \operatorname{arccosh}\!\left( \frac{2 r_{\infty}}{R} \right)
    \label{eq:rinf}
\end{equation}
We set $r_{\infty}$ to 40 Bohr, which is sufficient in most calculations.

To solve both the Sternheimer equation (Eq.~\ref{eq:radial st}) and the Poisson equation (Eq.~\ref{eq:Pos_epli}), we 
need to discretize the Laplace operator (Eq.~\ref{eq:Nabula_ellip}) in the prolate spheroidal coordinates system.
Eq.~\ref{eq:Nabula_ellip} involves only derivatives with respect to either $\mu$ or $\nu$ (no cross terms), which can be easily calculated using finite difference on a uniform grid. We employ the following 9-point central difference formulae for the first and second derivatives of a function $f(x)$,
\begin{equation}
    \begin{aligned}
f_{i}^{\prime}= & \frac{1}{840 h}\left(3 f_{i-4}-32 f_{i-3}+168 f_{i-2}-672 f_{i-1}\right. \\
& \left.+672 f_{i+1}-168 f_{i+2}+32 f_{i+3}-3 f_{i+4}\right)+O\left(h^{8}\right) \\
f_{i}^{\prime \prime}= & \frac{1}{5040 h^{2}}(-9 f_{i-4}+128 f_{i-3}-1008 f_{i-2}+\\
&8064 f_{i-1}-14350 f_{i} +8064 f_{i+1}-1008 f_{i+2}+\\
&128 f_{i+3}-9 f_{i+4})+O(h^{8})
\end{aligned}
\end{equation}
where $f_i=f(x(i))$ and $x$ can be either $\mu$ or $\nu$ variable.
Under finite difference, the result of the Laplace operator acting on a wave function on a certain grid point is related to the wave functions on its 16 surrounding grid points. The total number of grid points is usually tens of thousands. This means that the Laplacian matrix in the real space grids representation is very sparse, and we can transform the Sternheimer equation and Poisson equation to a general sparse linear equations, 
\begin{equation}
    Af=B\, 
\end{equation}
where $A$ is a sparse matrix with dimension $N_\mu N_{\nu}$, and the vast majority of non-diagonal elements in $A$ are 0.
We choose to call the \textit{pardiso} solver in the MKL library function, which can be used to solve large sparse matrix linear equation systems

\subsection{Scalar-relativistic Sternheimer equation within atomic ZORA approxiation.}
\label{appdenix_rel}
In FHI-aims, the scalar relativity is treated under the zeroth-order regular approximation (ZORA), where the kinetic energy operator is given by
\begin{equation}
    \hat{t}_{ZORA}=\bm{p}.\frac{c^2}{2c^2-v}.\bm{p}
    \label{eq:ZORA}
\end{equation}
Here, $\bm{p}$ represents the momentum operator,  $c$ the light speed, and $v$ the effective potential of the system.
To restore the gauge invariance against shifts of the potential zero in Eq.~\ref{eq:ZORA}, one can substitute in Eq.~\ref{eq:ZORA} for $v$ only the on-site free-atom potential $v^{free}_{at}(j)$ at the atomic center [$at(j)$] associated with a basis function $j$,
\begin{equation}
    \hat{t}_{at,ZORA}\left | \varphi_j \right \rangle =\bm{p}.\frac{c^2}{2c^2-v^{free}_{at(j)}}.\bm{p}\left | \varphi_j \right \rangle\, 
    \label{eq:at_ZORA}
\end{equation}
which is referred to as ``atomic ZORA". Obviously, this is a concept defined under atomic orbitals. The atom to which the basis function belongs will affect the kinetic energy operator itself. In this work, we solve the Sternheimer equation under the approximation of ``atomic ZORA” in real space grids 
to account for the scalar relativistic effect. For diatomic molecules,  it is natural and reasonable to define the kinetic energy operator 
on real space grids under atomic ZORA as follows,
\begin{equation}
    \hat{t}_{at,ZORA} f(\bm{r}) =\bm{p}.\frac{c^2}{2c^2-v^{free}_{near}(\bm{r})}.\bm{p} f(\bm{r})\, 
    \label{eq:at_ZORA_real_space}
\end{equation}
where ``near"  specifies the atom closer to a given spacial point $\bm{r}$. 

\section{Convergence test}
\subsection{Convergence with respect to the real-space grid density.}
Here, we test the convergence behavior of the all-electron (AE) RPA correlation energy with respect to the grid size of the real space for the dimers N$_2$, P$_2$ and As$_2$. The results obtained with increasing grid sizes are presented in Table~\ref{tab:grid_convergence}-\ref{tab:grid_As2} for the three dimers, respectively.
For the N$_2$ molecule, we tested the convergence of the RPA correlation energy with respect to the grid density under different $M_{\text{max}}$ and $N_{\text{eigen}}$ parameters. 
It can be seen that for all tested parameter sets, the $90\times90$ grid is already converged to the meV level, while the $150\times120$ grid reaches the 0.1~meV level. 
We also observe that as $M_{\text{max}}$ and $N_{\text{eigen}}$ increase, the error associated with a given grid density becomes larger. 
Comparing the smallest parameter set ($M_{\text{max}}=5$, $N_{\text{eigen}}=500$) with the largest one ($M_{\text{max}}=16$, $N_{\text{eigen}}=1500$), and taking the $192\times150$ grid as the reference, 
the errors of the $120\times100$ grid are 0.13~meV and 0.26~meV, respectively, whereas those of the $150\times120$ grid are 0.03~meV and 0.08~meV, respectively. 
This indicates that regardless of whether the smallest or largest parameter set is used, the corresponding grid errors remain of the same order of magnitude, 
and therefore do not affect the assessment of grid convergence. 
Consequently, for P$_2$ and As$_2$, it is sufficient to perform the grid-convergence test with a single reasonable parameter choice.
For heavier elements such as P$_2$ (see Table~\ref{tab:grid_P2}), grids approximately twice as dense as N$_2$ are needed to
achieve comparable accuracy. This trend continues for dimers formed with even heavier element; see the results for As$_2$ in Table~\ref{tab:grid_As2}. 

\begin{table}[htbp]
\centering
\caption{AE RPA@PBE for N$_{2}$  with different prolate spheroidal grid sizes under different  $M_{\text{max}}$ and $N_{\text{eigen}}$. 
The energy unit is eV.}
\label{tab:grid_convergence}
\begin{tabular}{ccccc}
\toprule
$M_{\text{max}}$ & $N_\mu \times N_\nu$ & $N_{\text{eigen}}=500$ & $N_{\text{eigen}}=1000$ & $N_{\text{eigen}}=1500$ \\
\midrule
\multirow{5}{*}{5} 
 & 70$\times$70   & -23.37226 & -23.37383 & -23.37410 \\
 & 90$\times$90   & -23.37172 & -23.37318 & -23.37351 \\
 & 120$\times$100 & -23.37147 & -23.37289 & -23.37319 \\
 & 150$\times$120 & -23.37137 & -23.37278 & -23.37307 \\
 & 192$\times$150 & -23.37134 & -23.37273 & -23.37301 \\
\midrule
\multirow{5}{*}{9} 
 & 70$\times$70   & -23.44688 & -23.44899 & -23.44938 \\
 & 90$\times$90   & -23.44625 & -23.44824 & -23.44871 \\
 & 120$\times$100 & -23.44596 & -23.44788 & -23.44832 \\
 & 150$\times$120 & -23.44583 & -23.44774 & -23.44816 \\
 & 192$\times$150 & -23.44579 & -23.44767 & -23.44808 \\
\midrule
\multirow{5}{*}{16} 
 & 70$\times$70   & -23.45606 & -23.45845 & -23.45889 \\
 & 90$\times$90   & -23.45541 & -23.45767 & -23.45823 \\
 & 120$\times$100 & -23.45510 & -23.45729 & -23.45781 \\
 & 150$\times$120 & -23.45497 & -23.45713 & -23.45763 \\
 & 192$\times$150 & -23.45492 & -23.45706 & -23.45755 \\
\bottomrule
\end{tabular}
\end{table}



\begin{table}[!h]
\caption{AE RPA@PBE for P$_{2}$  with different prolate spheroidal grid sizes. 
Here, $N_{\text{eigen}}=1000$ and $M_{max}=9$.  The energy unit is eV. }
       \label{tab:grid_P2}
\begin{tabular}{c c  }

\hline
$N_\mu \times N_\nu$ & $E_c^\text{RPA}(\text{P}_2$)     \\ \hline
120  $\times$ 100 & -50.27680 \\
150  $\times$ 120 & -50.27521 \\
200  $\times$ 150 &-50.27449\\
250  $\times$ 200 & -50.27431 \\

      \hline
      \hline
\end{tabular}
\end{table}

\begin{table}[!h]
\caption{AE RPA@PBE for As$_{2}$  with different prolate spheroidal grid sizes. Here, $N_{\text{eigen}}=500$ and $M_{\text{max}}=9$. Atomic ZORA is used for the relativistic correction. The energy unit is eV. }
\begin{tabular}{c c}

\hline
$N_\mu \times N_\nu$ & $E_c^\text{RPA}(\text{As}_2)$   \\ \hline
192 $\times$ 150 & -133.90221\\
240  $\times$ 180 &  -133.90095\\
288  $\times$ 200 &-133.90050\\
336  $\times$ 250 &-133.90017        \\

      \hline
      \hline
       \label{tab:grid_As2}
\end{tabular}
\end{table}

%
%

\subsection{Derivation of the $M_{\text{max}}$-dependence behavior of the RPA correlation energy} 
Here we derive the dependence of the RPA correlation energy on the magnetic quantum number $M_{\text{max}}$, i.e., Eq.~10 in the main text
for a diatomic system.
We assume that this dependence has the following form 
\begin{equation}
    E(M_{\text{max}}) = E_\infty + \frac{\alpha}{M_{\text{max}}^\beta}
    \label{eq:E_M_beta}
\end{equation}
where $E_\infty$ represents the correlation energy when $M_{\text{max}}$ tends towards infinity, and $\alpha$ is a parameter depending on the
atomic type and bond length. We assume that $\beta$ is independent of atomic types and bond length, as $M_{\text{max}}$ reflects the symmetry of diatomic molecules, which is not affected by atomic types and bond length. 

We first consider an isolated atom which can be regarded as a diatomic molecule with a large bond length. For an atom, one can analyze the convergence of its correlation energy both with the magnetic quantum number $M_{\text{max}}$, and with the azimuthal quantum number $L_{\text{max}}$. In fact, as will be
shown below, the dependence of $E_c^\text{RPA}$ on $M_{\text{max}}$, or more precisely the value of $\beta$ can be derived from the convergence behavior of the atomic RPA correlation energy with $L_{\text{max}}$.\\

Let's start from the convergence behavior of the atomic RPA correlation energy with $L_{\text{max}}$ \cite{peng2023basis} (In the following discussion, both $L_{\text{max}}$, $M_{\text{max}}$, $L$ and $M$ are large numbers), 
\begin{equation}
    \label{eq:atom_lmax}    
    E(L_{\text{max}}) = E_\infty + \frac{\gamma}{L_{\text{max}}^3}\, 
\end{equation}
The energy contribution of a given angular momentum channel $L$ can be obtained as
\begin{equation}
\begin{aligned}
     \Delta E(L)&= E(L)-E(L-1)\\
     &=\gamma(L^{-3}-(L-1)^{-3}) \\
     &\approx -\frac{3\gamma}{L^4} \\
     &=\frac{\gamma^\prime}{L^4} \, .
\end{aligned}
\end{equation}
This is also well known from early quantum chemistry studies \cite{Schwartz:1962,Hill:1985}.
One can obtain the energy contribution $\Delta E(M)$, i.e., the correlation energy contribution from a given
magnetic momentum channel $M$, from the $\Delta E(L)$,
\begin{equation}
\begin{aligned}
     \Delta E(M)&= \sum_{L=M}^\infty \frac{2\Delta E(L)}{2L+1}\, .
\end{aligned}
\end{equation}
Here, $2L+1$ term  means  $2L+1$ degenerate  channels under the $L$ channel, and  $2$ in the numerator means that $+M$ 
and $-M$ contribute equally to the correlation energy and both belong to $E(M)$. Thus, for large $M$, one can get,
\begin{equation}
\begin{aligned}
     \Delta E(M)&= \sum_{L=M}^\infty \frac{2\gamma^\prime}{(2L+1)L^4}\\
    &\approx \sum_{L=M}^\infty \frac{\gamma^\prime}{L^5}\\
     &\approx \int_{L=M}^\infty dL \gamma^\prime L^{-5}\\
     &=\frac{\gamma^\prime}{4}M^{-4}
\end{aligned}
\end{equation}
It can be easily seen that $\Delta E(M)$ and $\Delta E(L)$ have the same form, except that the coefficients are different.
Therefore, $E(M)$ and $E(L)$ also have the same asymptotic behavior, and one can be assured that $\beta=3$ in Eq.~\ref{eq:E_M_beta}.

\begin{figure}[htbp]
    \centering
   \includegraphics[scale=0.3]{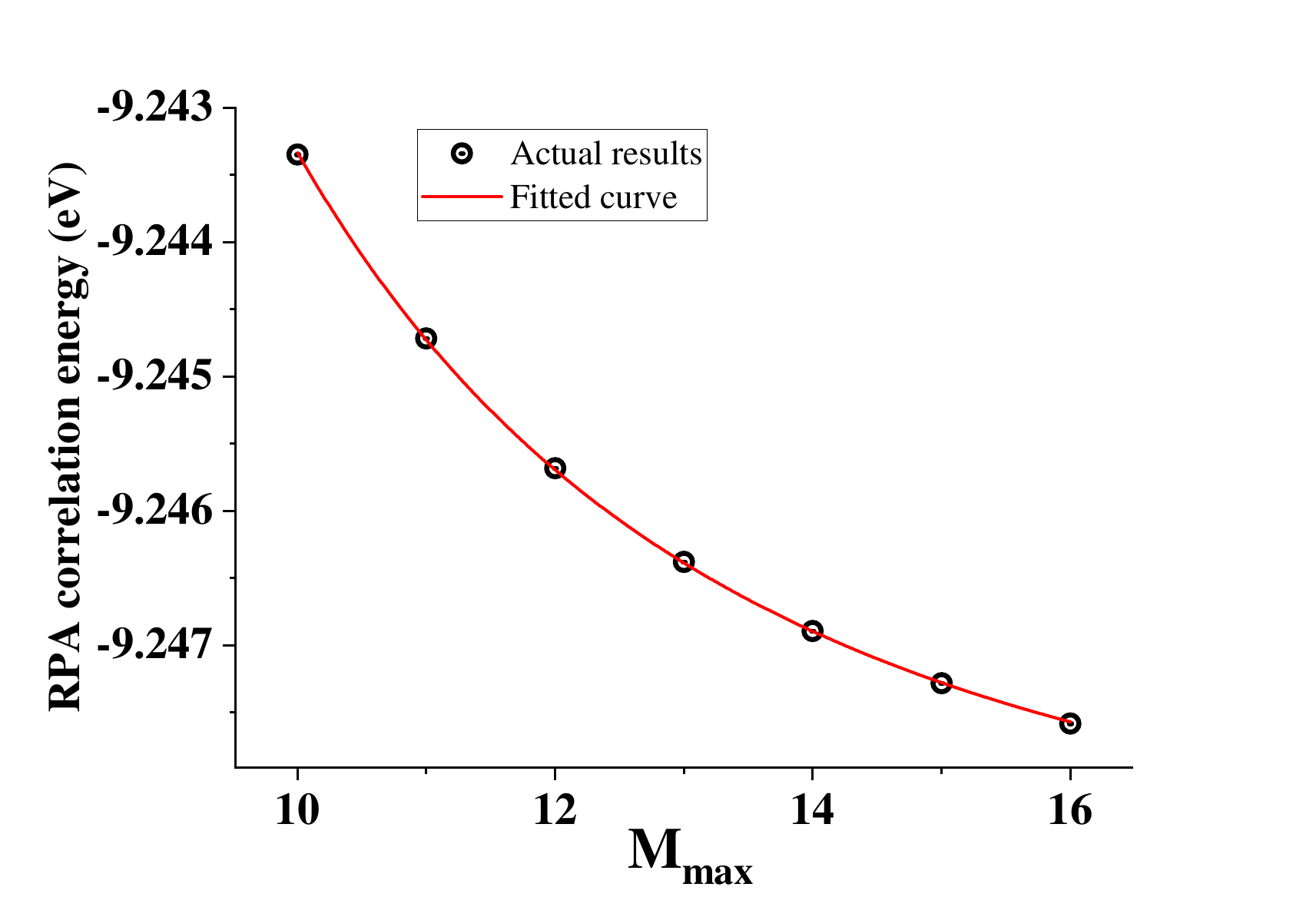}
    \caption{\label{fig:N_mmax_fit} The convergence behavior of correlation energy with respect to $M_{\text{max}}$ for the N atom }    
\end{figure}

\begin{figure}[htbp]    
    \centering
    \includegraphics[scale=0.3]{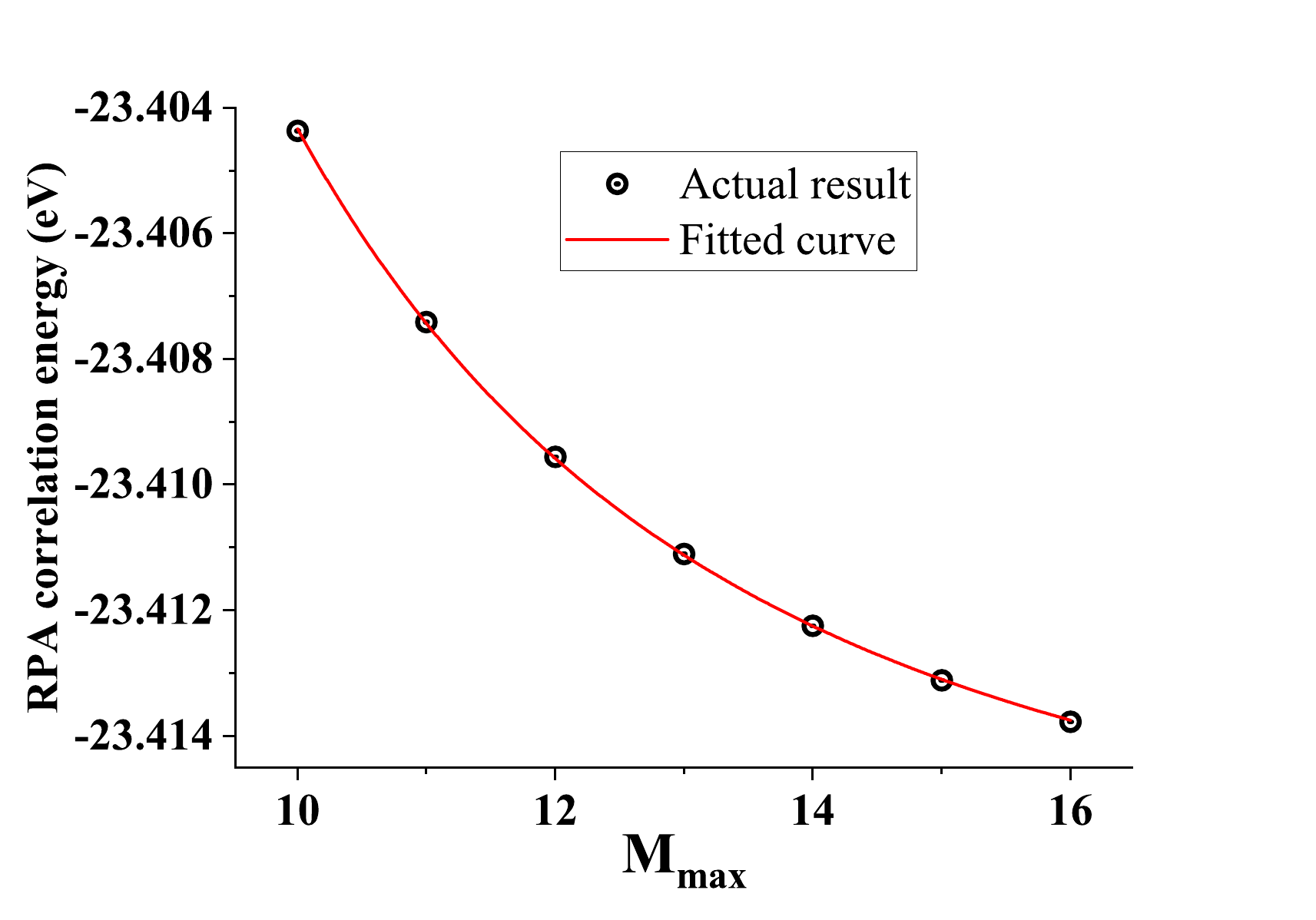}
    \caption{\label{fig:N2_mmax_fit} The convergence behavior of correlation energy with respect to $M_{\text{max}}$ for N$_2$. }   
\end{figure}

To verify this numerically, we plot $E_c^\text{RPA}(M_{\text{max}})$ as 
a function of $M_{\text{max}}$ for the N atom in Fig.~\ref{fig:N2_mmax_fit}, and then
fitted the data from $M_{\text{max}}=10$ to $M_{\text{max}}=16$ using Eq.~\ref{eq:E_M_beta} with $\beta=3$. 
The confidence level $R^2$ of the fitting equals 0.99995, indicating that the calculated data is highly consistent with
the fitting expression, as can also be seen from Fig.~\ref{fig:N_mmax_fit}.
A fitting for the results of the N atom
yields $E_c^{\infty}(\infty)=-9.24857$, and $\alpha=5.52887$. 


We expect that the correlation energy of any system with cylindrical symmetry has the same asymptotic dependence on $M_{\text{max}}$.
Thus, Eq.~\ref{eq:E_M_beta} with $\beta=3$ should apply to any diatomic molecule (the isolated atom is a special case with infinite bond length).
To verify if this is really the case, we plot $E_c^\text{RPA}(M_{\text{max}})$ as 
a function of $M_{\text{max}}$ for N$_2$ in Fig.~\ref{fig:N2_mmax_fit}, and then
fitted the data from $M_{\text{max}}=10$ to $M_{\text{max}}=16$ using Eq.~\ref{eq:E_M_beta} with $\beta=3$. 
The confidence level $R^2$ of the fitting equals 0.99995, indicating that the calculated data is highly consistent with
the fitting expression, as can also be seen from Fig.~\ref{fig:N2_mmax_fit}. Via the fitting procedure, we obtained
$E_c^\text{RPA}(\infty)=-23.41679$ eV for the N$_2$ molecule, with the coefficient $\alpha=12.45267$ eV.

To further verify the convergence behavior of the RPA correlation energy with respect to 
$1/M_{\text{max}}^{3}$, we increased $M_{\text{max}}$ from 16 to 30. 
The absolute molecular energies at different angular momentum cutoffs are summarized in 
Table~\ref{tab:rpa_convergence}, where the directly computed results are compared with the fitted ones. 
The results are as follows:

\begin{table}[htbp]
\centering
\caption{{Comparison of RPA-calculated and fitted absolute molecular energies (in eV) 
for different $M_{\text{max}}$. The second column of the table represents the directly computed results. 
The third and fourth columns correspond to the fitted results obtained 
using the data at $M_{\text{max}} = 10, 11, 12$ and 
$M_{\text{max}} = 14, 15, 16$, respectively. The values in parentheses represent the deviations from the actual calculated results (first column), in units of meV. The last line reports the RPA correlation energy extrapolated to $M_\text{max} = \infty$.}}
\label{tab:rpa_convergence}
\begin{tabular}{cccc}
\hline
$M_{\text{max}}$ & RPA-Calculate & RPA-Fitting (10--12) & RPA-Fitting (14--16) \\
\hline
0  & -12.32344 &             &             \\
1  & -20.73323 &             &             \\
2  & -22.58044 &             &             \\
3  & -23.09092 &             &             \\
4  & -23.25902 &             &             \\
5  & -23.32929 &             &             \\
6  & -23.36355 &             &             \\
7  & -23.38215 &             &             \\
8  & -23.39307 &             &             \\
9  & -23.39990 &             &             \\
10 & -23.40437 &             &             \\
11 & -23.40741 &             &             \\
12 & -23.40956 &             &             \\
13 & -23.41110 &             &             \\
14 & -23.41225 &             &             \\
15 & -23.41311 &             &             \\
16 & -23.41377 &             &             \\
17 & -23.41429 & -23.41416(0.13) & -23.41429(0.00) \\
18 & -23.41470 & -23.41456(0.16) & -23.41470(0.00) \\
19 & -23.41502 & -23.41488(0.16) & -23.41502(0.00) \\
20 & -23.41529 & -23.41513(0.16) & -23.41528(0.01) \\
21 & -23.41550 & -23.41534(0.16) & -23.41550(0.00) \\
22 & -23.41567 & -23.41551(0.16) & -23.41568(-0.01) \\
23 & -23.41582 & -23.41566(0.16) & -23.41583(-0.01) \\
24 & -23.41594 & -23.41578(0.16) & -23.41595(-0.01) \\
25 & -23.41604 & -23.41588(0.16) & -23.41606(-0.02) \\
26 & -23.41613 & -23.41597(0.16) & -23.41615(-0.02) \\
27 & -23.41620 & -23.41604(0.16) & -23.41623(-0.03) \\
28 & -23.41627 & -23.41611(0.16) & -23.41629(-0.02) \\
29 & -23.41632 & -23.41617(0.15) & -23.41635(-0.03) \\
30 & -23.41637 & -23.41621(0.16) & -23.41640(-0.03) \\
\hline
$\infty$  &   / &-23.41667 &-23.41687 \\
\hline
\end{tabular}
\end{table}
It can be seen that the results fitted using $M_{\text{max}}=10,11,12$ differ from the directly computed values by about $0.16$~meV. 
For example, at $M_{\text{max}}=20,25,30$, the absolute error amounts to $0.16$~meV. 
Moreover, when the fitting is performed with $M_{\text{max}}=14,15,16$, the errors at 
$M_{\text{max}}=20,25,30$ are reduced to $0.01$~meV, $0.02$~meV, and $0.03$~meV, respectively. 
This clearly demonstrates the high consistency between the fitted and directly computed results. 
It is easy to understand that fittings based on data with higher angular momentum cutoffs perform better, 
since the $1/M_{\text{max}}^{3}$ behavior only holds in the asymptotic regime of large $M_{\text{max}}$. 
Therefore, the above analysis shows that the extrapolation with respect to $M_{\text{max}}$ in this work 
is fundamentally different from the conventional basis-set extrapolation of correlation energies. 
Both the fitting formula and the required input data are highly accurate. 
Thus, the role of fitting in the present work is merely to accelerate the convergence, rather than being indispensable. 
In principle, one can always achieve the desired accuracy by systematically increasing the angular momentum cutoff, 
without relying on any fitting procedure.

\subsection{Convergence with respect to the $r_{\infty}$.}
Here we examine the influence of the choice of the infinite cutoff radius $r_{\infty}$ on the numerical results. 
We tested the AE RPA correlation energy of  N$_2$ and P$_2$ molecules by varying $r_{\infty}$ and monitoring the corresponding changes in the energy. 
Since increasing $r_{\infty}$ leads to a sparser grid distribution, 
we also performed calculations with different grid sizes. 
In these tests, the bond lengths of N$_2$ and P$_2$ were set to 1.104~\AA{} and 1.907~\AA{}, respectively, 
using 32 minimax grid points with $M_{\text{max}}=9$ and $N_{\text{eigen}}=1000$.
It should be noted that the effect of $r_{\infty}$ on the grid density is expected to be weak, because, according to Eq.~\ref{eq:rinf}, $r_{\infty}$ is not linearly related to the actual grid spacing $u_{\infty}$ used in the finite-difference scheme. Increasing $r_{\infty}$ mainly affects the spacing of grid points near infinity, without significantly influencing the distribution of points near the nuclei, which is confirmed by the actual calculations.
\begin{table}[htbp]
\centering
\caption{Dependence of the total energy (eV) of N$_2$ on the cutoff radius $r_{\infty}$ (a.u.) with different grid sizes.}
\label{tab:rmax_n2}
\begin{tabular}{ccccc}
\toprule
$N_\mu \times N_\nu$ & $r_{\infty}=40$ & $r_{\infty}=60$ & $r_{\infty}=80$ &$r_{\infty}=100$ \\
\midrule
150$\times$120 & -23.44327 & -23.44402 & -23.44422 & -23.44430 \\
192$\times$150 & -23.44321 & -23.44394 & -23.44413 & -23.44420 \\
250$\times$200 & -23.44319 & -23.44392 & -23.44410 & -23.44417 \\
\bottomrule
\end{tabular}
\end{table}

\begin{table}[htbp]
\centering
\caption{Dependence of the total energy (eV) of P$_2$ on the cutoff radius $r_{\infty}$ (a.u.) with different grid sizes.}
\label{tab:rmax_p2}

\begin{tabular}{ccccc}
\toprule
$N_\mu \times N_\nu$ & $r_{\infty}=40$ & $r_{\infty}=60$ & $r_{\infty}=80$ &$r_{\infty}=100$ \\
\midrule
150$\times$120 & -50.2806 & -50.28373 & -50.28470 & -50.28519  \\
192$\times$150 & -50.2798 & -50.28266 & -50.28343 & -50.28377  \\
250$\times$200 & -50.2796 & -50.28241 & -50.28310 & -50.28335  \\
\bottomrule
\end{tabular}
\end{table}

For the N$_2$ molecule, with a grid size of $250 \times 200$, increasing $r_{\infty}$ from 80~Bohr to 100~Bohr changes the total energy by only 0.07~meV. From 40~Bohr to 100~Bohr, the energy changes by approximately 1~meV. Moreover, the energy differences for different grid sizes remain small.

 For P$_2$, with a grid size of $250 \times 200$, increasing $r_{\infty}$ from 80~Bohr to 100~Bohr results in an energy change of 0.25~meV. From 40~Bohr to 100~Bohr, the energy varies by about 3.7~meV. Again, the grid size has little impact on this conclusion.

Next, we tested the effect of $r_{\infty}$ on the absolute energy and binding energy, where molecules containing different elements were included, the result is showed in Tab.~\ref{tab:molecule_rmax}. Here the binding energy refers to the RPA correlation contribution to the binding, i.e., the RPA correlation energy of the molecule minus the sum of the RPA correlation energies of the atoms. It can be seen that, at the absolute RPA correlation energy level, $r_\infty=40$ Bohr introduces an error of less than approximately 4 meV. $\Delta E_B$ is much smaller than $\Delta E$, and only for P$_2$ does the binding energy error exceed 1 meV. This indicates that although the distant spatial region affects the absolute energy, it does not necessarily contribute to the binding energy.

\begin{table*}[htbp]
\centering
\setlength{\tabcolsep}{6pt} 
\caption{Total energies and binding energies of selected molecules at different cutoff radii ($r_{\infty}$ in Bohr). $E$ and $E_B$ represent the absolute RPA correlation energy and the binding energy contribution from the RPA correlation energy, respectively, and the numbers in parenthesis indicate the values of $r_{\infty}$. $\Delta$ indicates the difference between $r_{\infty}=100$ and $r_{\infty}=40$ Bohr. Energies are in eV (The units of $\Delta E$ and $ \Delta E_B$ are meV)  . For N$_2$ and O$_2$, the grid size is 192$\times$150; for P$_2$, Cl$_2$, GeO grid size is 250$\times$200. The scalar-relativistic atomic ZORA approximation is used for GeO.} 
\label{tab:molecule_rmax}
\begin{tabular}{lcccccc}
\toprule
MO & $E(40)$ & $E(100)$ & $\Delta E$ (meV) & $E_B(40)$ & $E_B(100)$ & $\Delta E_B$ (meV) \\
\midrule
N$_2$  & -23.44321 & -23.44420 & 0.99 & -4.95946 & -4.95981 & 0.35 \\
P$_2$  & -50.27963 & -50.28335 & 3.72 & -3.70426 & -3.70578 & 1.52 \\
O$_2$  & -27.46251 & -27.46334 & 0.83 & -3.82946 & -3.82979 & 0.33 \\
Cl$_2$ & -57.88060 & -57.88298 & 2.38 & -1.48542 & -1.48601 & 0.59 \\
GeO    & -80.16672 & -80.16900 & 2.28 & -4.78606 & -4.78666 & 0.60 \\
\bottomrule
\end{tabular}
\end{table*}

\subsection{Convergence with respect to the occupied states given by numerical atomic orbital.}
In this work, the occupied states are obtained based on calculations with
atomic orbital basis sets. In principle, one could iteratively solve the
DFT Hamiltonian in an ellipsoidal coordinate system to obtain occupied
orbitals and energies that are completely numerical and independent of
any basis set. However, performing DFT (or HF) level calculations in
ellipsoidal coordinates is neither the focus of the present study nor an
unsolved problem, as there already exist a considerable number of
published works in this area. On the other hand, the built-in \texttt{tier}
basis sets of \textsc{FHI-aims} are well capable of describing the occupied
states of molecules. Therefore, we believe that starting from occupied
states obtained with numerical atomic orbital basis sets will not
introduce errors beyond the meV level. The following numerical tests
confirm this point. We first examined the N$_2$ and P$_2$ molecules. Using
\texttt{tier2}, \texttt{tier3}, \texttt{tier4}, and \texttt{tier4} plus further basis
functions (hereafter denoted as \texttt{tier4+}), we investigated the
convergence of the DFT total energy, the 
(non-self-consistent) Hartree-Fock energy (evaluated using PBE orbitals, including exact exchange (EXX) energy, kinetic energy, Hartree energy and external potential energy), the RPA correlation energy obtained in this work. The inclusion of tests on the basis-set convergence of the total DFT and HF energies here is because both are determined solely by the occupied states, and thus can also reflect the convergence of the occupied states with respect to the basis set.    At the level of binding energies, we tested the basis-set convergence of the (non-self-consistent) Hartree-Fock contribution and the RPA correlation contribution to the binding energies.
For comparison, we also
computed results using the Gaussian basis sets aug-cc-pwCVXZ. The bond
lengths of N$_2$ and P$_2$ are 1.104~\AA{} and 1.907~\AA{}, respectively. The
parameters used for the RPA calculations were:
$r_{\infty} = 100$~Bohr, $N_{\text{eigen}} = 1000$, $M_{\text{max}} = 9$, with 32 minimax
frequency grid points (for N$_2$) and 96 Gauss--Legendre grid points (for P$_2$). From Table.~\ref{tab:ab-occ}, 
it can be seen that at the level of absolute energies, the total DFT and HF energies converge faster with the NAO basis sets than with the GTO basis sets. By examining the energy differences between basis set 4 and basis set 5, one observes that for the NAO basis sets, the energies change by only a few meV, whereas for the GTO basis sets, the change is an order of magnitude larger. For the RPA correlation energies, this trend is less pronounced. Regardless of whether NAO or GTO is used, the energy differences from basis set 4 to 5 are less than 1 meV. When the largest basis set (basis set 5) is employed, the NAO and GTO results match very well, with energy differences on the meV scale for DFT total energy and non-self-consistent HF energy, and sub-meV for RPA correlation energy.

From Table.~\ref{tab:bind-occ}, at the level of binding energies, the GTO basis sets exhibit slightly better convergence for the HF contribution, while the RPA contribution shows no significant difference. Based on the results of basis set 5, the binding energies obtained with NAO and GTO are very close for both HF (meV level difference ) and RPA (sub-meV level difference) contributions. The total binding energies differ by only 0.39 meV for N$_2$ and 1.73 meV for P$_2$.
\begin{table*}[htbp]
    \centering
    \caption{Converge test of DFT, HF, and RPA results with NAO and GTO basis sets for N$_2$ and P$_2$. Basis set 2,3,4,5 represents the tier2,tier3,tier4,tier4+ (for NAO) and aug-cc-pwCVDZ, aug-cc-pwCVTZ, aug-cc-pwCVQZ, aug-cc-pwCV5Z (for GTO). Energy unit is eV.}
    \footnotesize
    \begin{tabular}{c|c|cc|cc|cc}
        \hline
        \multirow{2}{*}{Molecule} & \multirow{2}{*}{Basis-set} 
        & \multicolumn{2}{c|}{DFT} & \multicolumn{2}{c|}{HF} & \multicolumn{2}{c}{RPA} \\
        \cline{3-8}
         &  & NAO & GTO & NAO & GTO & NAO & GTO \\
        \hline
        \multirow{4}{*}{N$_2$}
         & 2 & -2978.53785 & -2977.69764 & -2965.25932 & -2964.51162 & -23.43728 & -23.43926 \\
         & 3 & -2978.55721 & -2978.34691 & -2965.28609 & -2965.09353 & -23.44307 & -23.44436 \\
         & 4 & -2978.56261 & -2978.52635 & -2965.28077 & -2965.24746 & -23.44436 & -23.44348 \\
         & 5 & -2978.56450 & -2978.56063 & -2965.28363 & -2965.28040 & -23.44420 & -23.44410 \\
        \hline
        \multirow{4}{*}{P$_2$}
         & 2 & -18569.73276
 & -18568.91365 &-18543.70282
 & -18543.30497 & -50.31838 
 & -50.25893 \\
         & 3 &-18569.7398
 & -18569.52701 & -18543.71209
 & -18543.62340 & -50.31863 
 & -50.31603 \\
         & 4 & -18569.74225
 & -18569.71004 & -18543.72133
 & -18543.71231 & -50.31700 
 
 & -50.31531 \\
         & 5 &-18569.74211
 & -18569.73504 & -18543.72281
 & -18543.72108 & -50.31631 
 & -50.31602 \\
        \hline
    \end{tabular}
    \label{tab:ab-occ}
\end{table*}

\begin{table*}[htbp]
    \centering
    \caption{Converge test of RPA binding ernergy  with NAO and GTO basis sets for N$_2$ and P$_2$. HF represents the binding energy contribution from the non-self-consistent Hartree-Fock part, and RPA represents the binding energy contribution from the RPA correlation part. Basis set 2,3,4,5 represents the tier2,tier3,tier4,tier4+ (for NAO) and aug-cc-pwCVDZ, aug-cc-pwCVTZ, aug-cc-pwCVQZ, aug-cc-pwCV5Z (for GTO). Energy unit is eV.}
    \footnotesize
    \begin{tabular}{c|c|cc|cc|cc}
        \hline
        \multirow{2}{*}{Molecule} & \multirow{2}{*}{Basis-set} 
        & \multicolumn{2}{c|}{HF} & \multicolumn{2}{c|}{RPA} & \multicolumn{2}{c}{Total} \\
        \cline{3-8}
         &  & NAO & GTO & NAO & GTO & NAO & GTO \\
        \hline
        \multirow{4}{*}{N$_2$}
         & 2 & -4.79231 & -4.51980 & -4.94263 & -4.97085 & -9.73494 & -9.49065 \\
         & 3 & -4.79022 & -4.75177 & -4.95719 & -4.95823 & -9.74741 & -9.71000 \\
         & 4 & -4.76971 & -4.77205 & -4.95965 & -4.95852 & -9.72936 & -9.73058 \\
         & 5 & -4.77208 & -4.77311 & -4.95981 & -4.95918 & -9.73189 & -9.73228 \\
        \hline
        \multirow{4}{*}{P$_2$}
         & 2 & -1.39475
 & -1.21177 &-3.69917 
 & -3.72506 & -5.09254 & -4.93683 \\
         & 3 & -1.39855
 & -1.36463 &-3.69956 
 & -3.70406 & -5.09707 & -5.06868 \\
         & 4 & -1.38765
 & -1.38032 & -3.70229 
 & -3.70377 & -5.08927 & -5.08409 \\
         & 5 & -1.38177
 & -1.38009 & -3.70390
 & -3.70385 & -5.08567 
 & -5.08394 \\
        \hline
    \end{tabular}
    \label{tab:bind-occ}
\end{table*}

Therefore, the following conclusion can be drawn. In this work, the calculation of RPA correlation energies exhibits excellent convergence with respect to the occupied states in the LCAO framework. Both the absolute RPA correlation energies and the RPA correlation contributions to binding energies converge to at least the meV level. When using the largest NAO or GTO basis sets, the absolute RPA correlation energies obtained from the two types of basis sets agree to within sub-meV accuracy. In Tab.~\ref{tab:molecule_tiers} and Tab.~\ref{tab:molecule_tiers2}, we tested additional diatomic molecules composed of various periodic elements, examining the convergence of both their absolute RPA correlation energies and the RPA contributions to binding energies with respect to the NAO basis sets. The results show that, for both absolute RPA correlation energies and binding energies, the NAO basis sets converge to at least the meV level.

\begin{table}[htbp]
\centering
\scriptsize
\caption{Converge test of RPA correlation ernergy for selected
molecules under NAO basis set. The energy unit is eV. Since chlorine does not have a tier4 basis set, for Cl$_2$ the last column represents the results obtained using tier3 plus futher basis functions.}
\label{tab:molecule_tiers}
\begin{tabular}{|c|c|c|c|c|}
\hline
Molecule & tier2 & tier3 & tier4 & tier4+ \\
\hline
N$_2$   & -23.43728 & -23.44307 & -23.44436 & -23.44420 \\
P$_2$   & -50.31838 & -50.31863 & -50.31700 & -50.31631 \\
O$_2$   & -27.47257 & -27.46522 & -27.46414 & -27.46334 \\
Cl$_2$  & -57.88487 & -57.88348 & --- & -57.88298 \\
GeO     & -80.17571 & -80.17090 & -80.16894 & -80.16900 \\
\hline
\end{tabular}
\end{table}

\begin{table}[htbp]
\centering
\scriptsize
\caption{Converge test of binding energy (contributed by RPA correlation energy) for selected
molecules under NAO basis set. The energy unit is eV. Since chlorine does not have a tier4 basis set, for Cl$_2$ the last column represents the results obtained using tier3 plus futher basis functions.}
\label{tab:molecule_tiers2}
\begin{tabular}{|c|c|c|c|c|}
\hline
Molecule & tier2 & tier3 & tier4 & tier4+ \\
\hline
N$_2$   & -4.94263 & -4.95719 & -4.95965 & -4.95981 \\
P$_2$   & -3.69917
 & -3.69956 & -3.70229 & -3.70390 \\
O$_2$   & -3.82817 & -3.82700 & -3.82987 & -3.82979 \\
Cl$_2$  & -1.47775 & -1.48461 & ---      & -1.48542 \\
GeO     & -4.79230 & -4.78631 & -4.78628 & -4.78666 \\
\hline
\end{tabular}
\end{table}

\section{Assessing the basis set superposition error}
 \begin{figure}[htbp]  
    \includegraphics[scale=0.32]{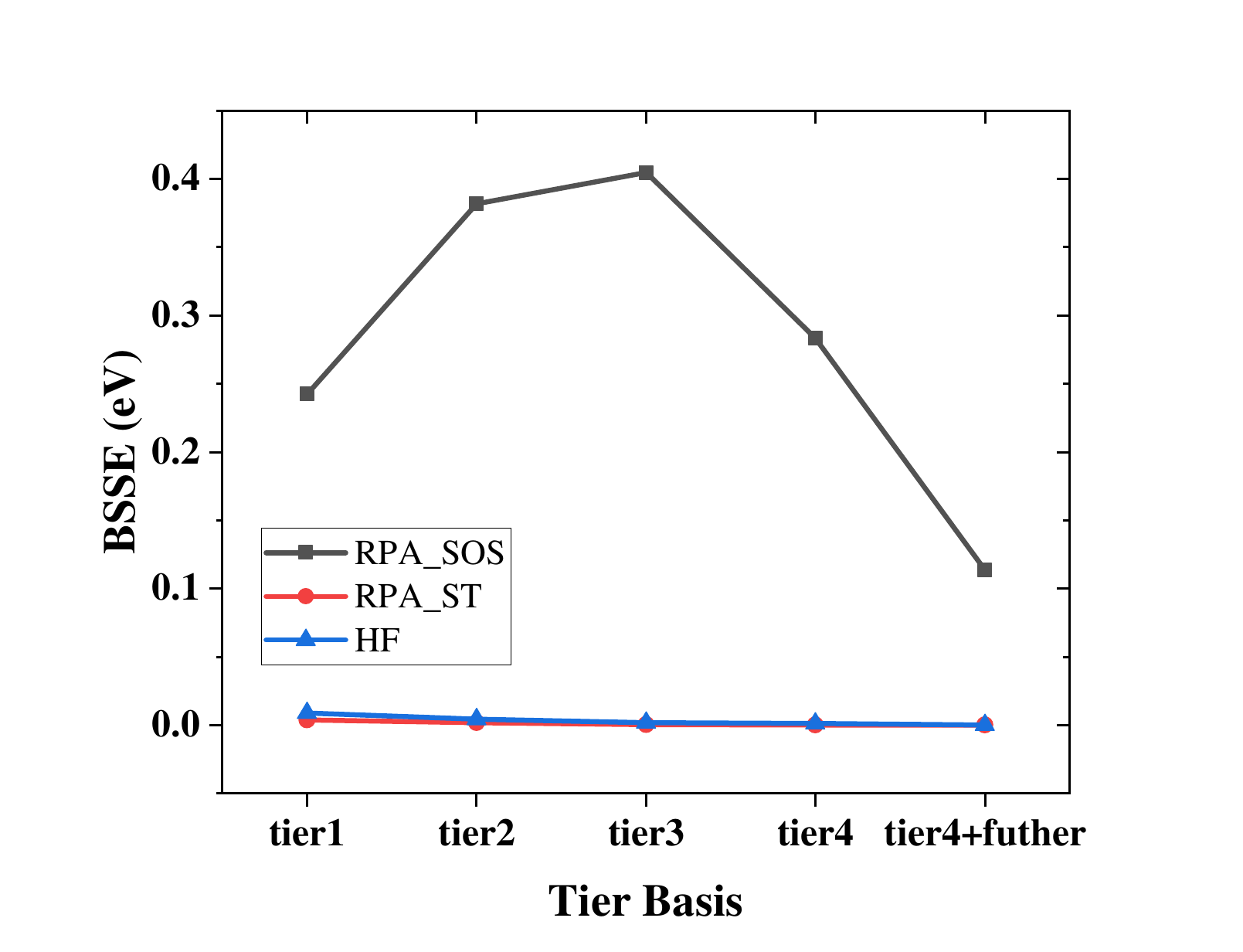}
    \includegraphics[scale=0.32]{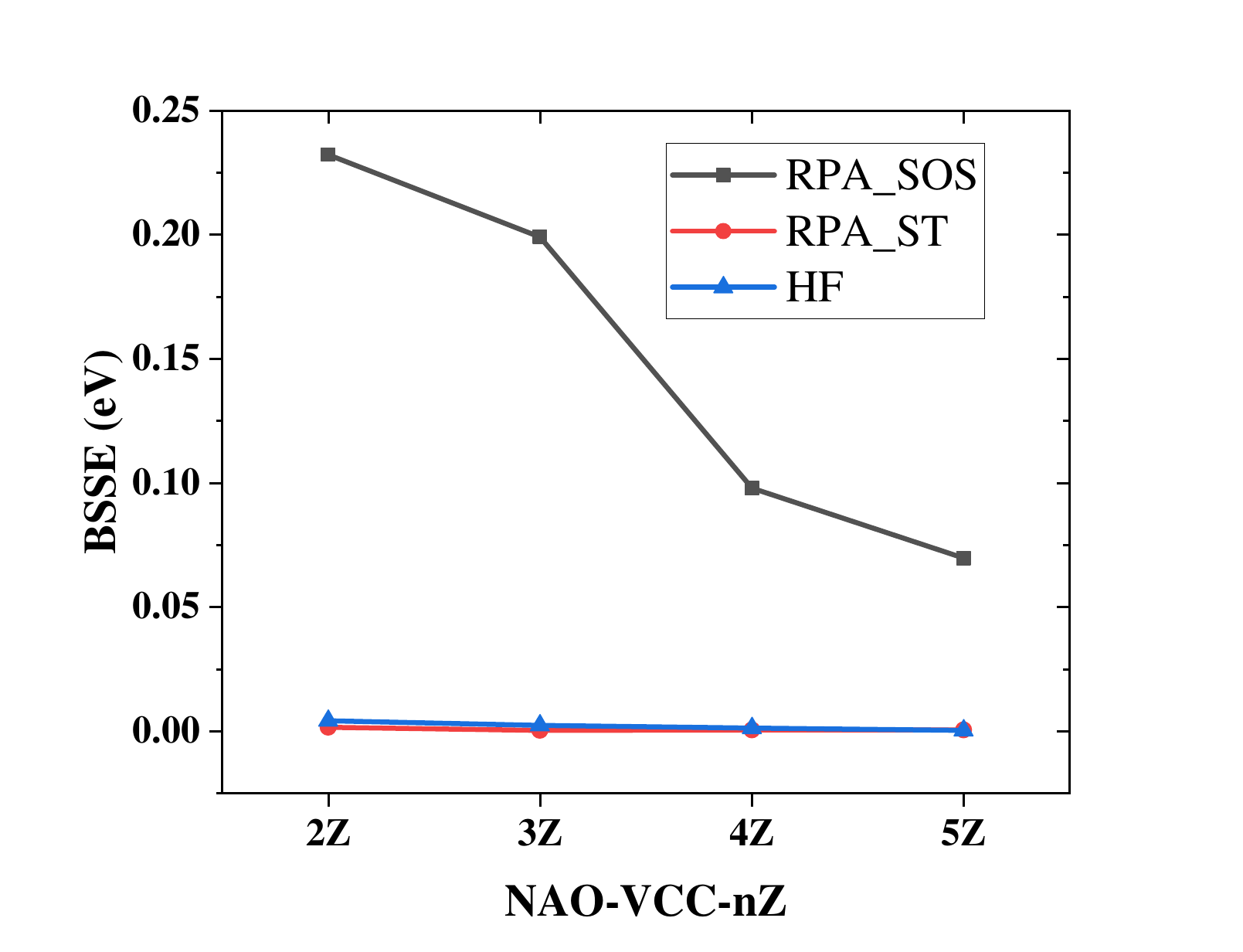}
    \includegraphics[scale=0.32]{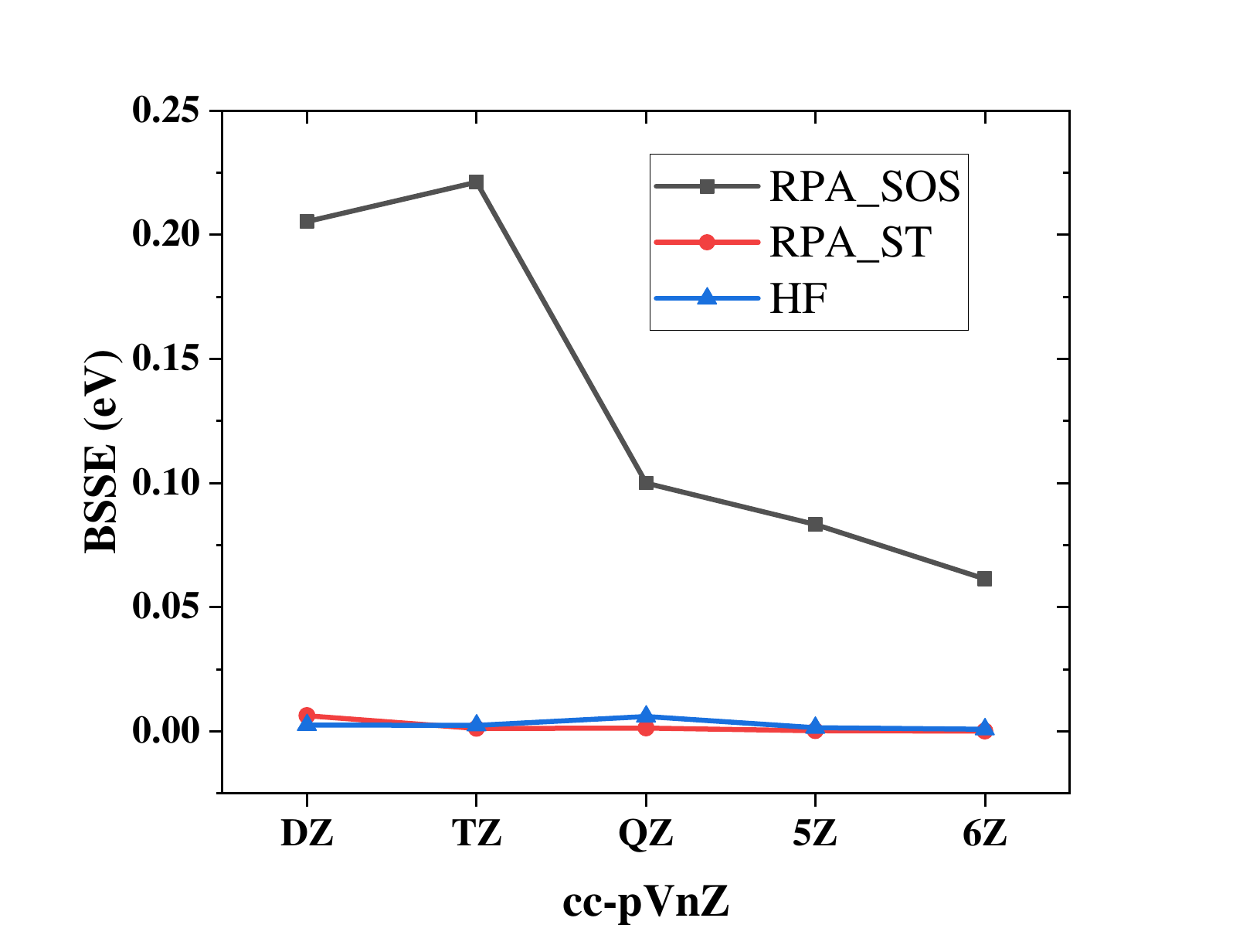}
    \caption{\label{fig:BSSE} BSSEs for the N$_2$ dimer for three series of AO basis sets. The value of BSSE is obtained by subtracting the energy of an atom with a ghost atom around from that of an isolated atom. The black line represents the BSSE of the RPA correlation energy obtained using the traditional SOS method, while the red line represents the BSSE of the RPA correlation energy obtained based on the present method. The blue line represents the BSSE error of the (non-self-consistent) Hartree-Fock energy component.  }
\end{figure}
Here we compare the basis set superposition error (BSSE) present
in the traditional ``sum over states (SOS)" scheme and the present Sternheimer
approach for three series of atomic orbital (AO) basis sets. Again, the N$_2$ molecule is used
as an illustration example. We estimate the BSSE using the Boys-Bernardi counterpoise scheme \cite{Boys/Bernardi:1970}, where
the BSSE is given by the difference between
the energy of an atom with a ghost atom nearby and that of an isolated atom.
The three panels in Table~\ref{fig:BSSE} present the BSSEs for three types of AO basis sets, including two series of NAOs (``tiers" and NAO-VCC-$n$Z) and one series of GTOs (cc-pV$n$Z). In each panel, we plot the BSSEs of the RPA correlation energy obtained with the traditional SOS scheme and the
present Sternheimer scheme,as well as the (non-self-consistent) Hartree-Fock energy (including exact exchange (EXX) energy, kinetic energy, hartree energy and externel potential energy). 
As is evident, in the traditional approach, the BSSEs are huge and do not vanish
even with the largest available basis set in each series.
Such BSSEs mainly come from the imbalanced description of the unoccupied manifold of the Hilbert space of
the molecule and the atom. In contrast, within the present approach, the BSSEs are vanishingly small, and these tiny
BSSEs come from the BSIE for the AO basis sets for
describing the occupied manifold of the Hilbert space. Hence, the magnitude of the BSSEs in the present Sternheimer
approach is similar to that of the Hartree-Fock part of the total energy.
Apparently, in our approach, there is no need to invoke
the counterpoise procedure to correct BSSEs any more.


\section{Error due to the frozen-core approximation}

\begin{figure}[htbp]   
    \centering
    \includegraphics[scale=0.35]{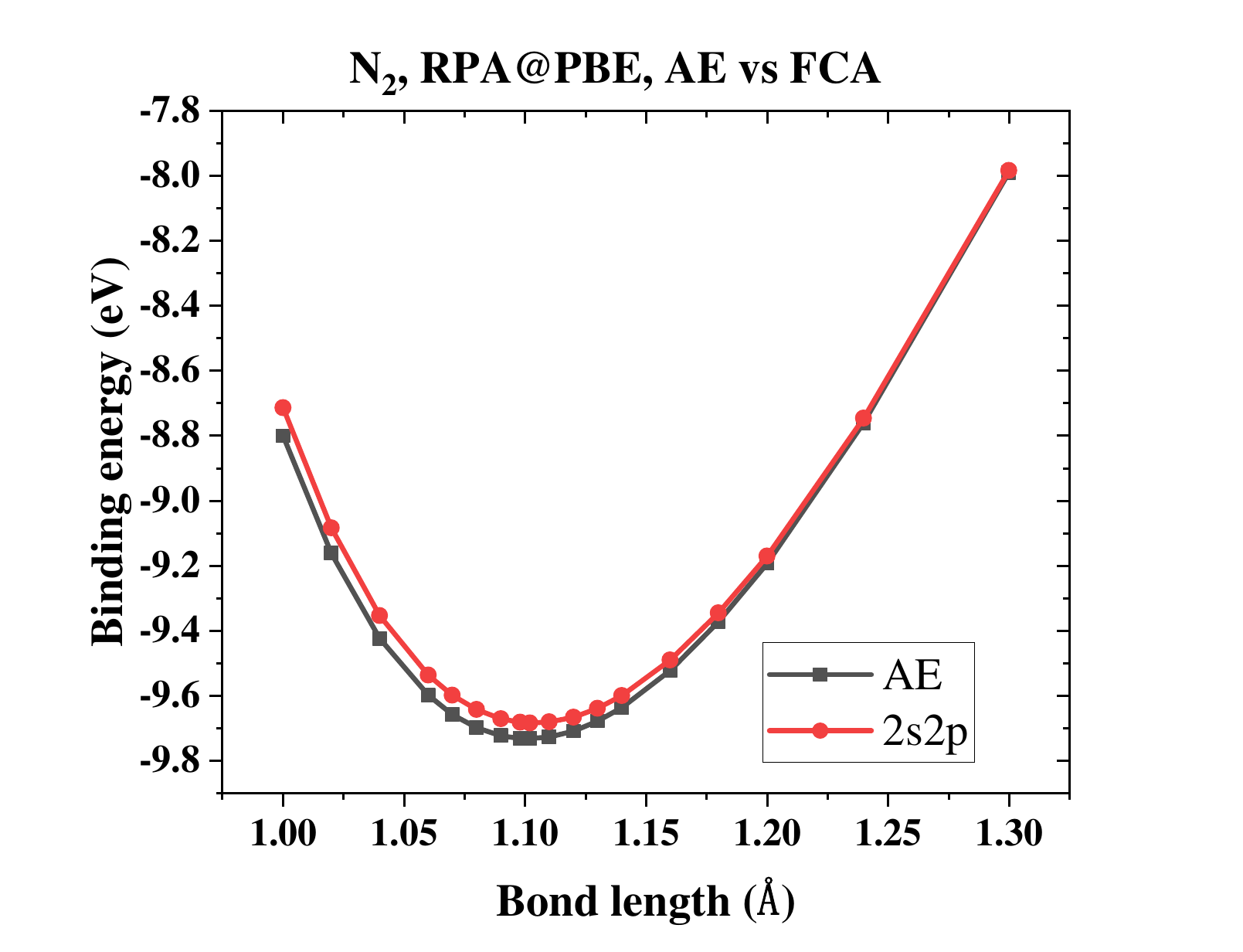}
    \includegraphics[scale=0.35]{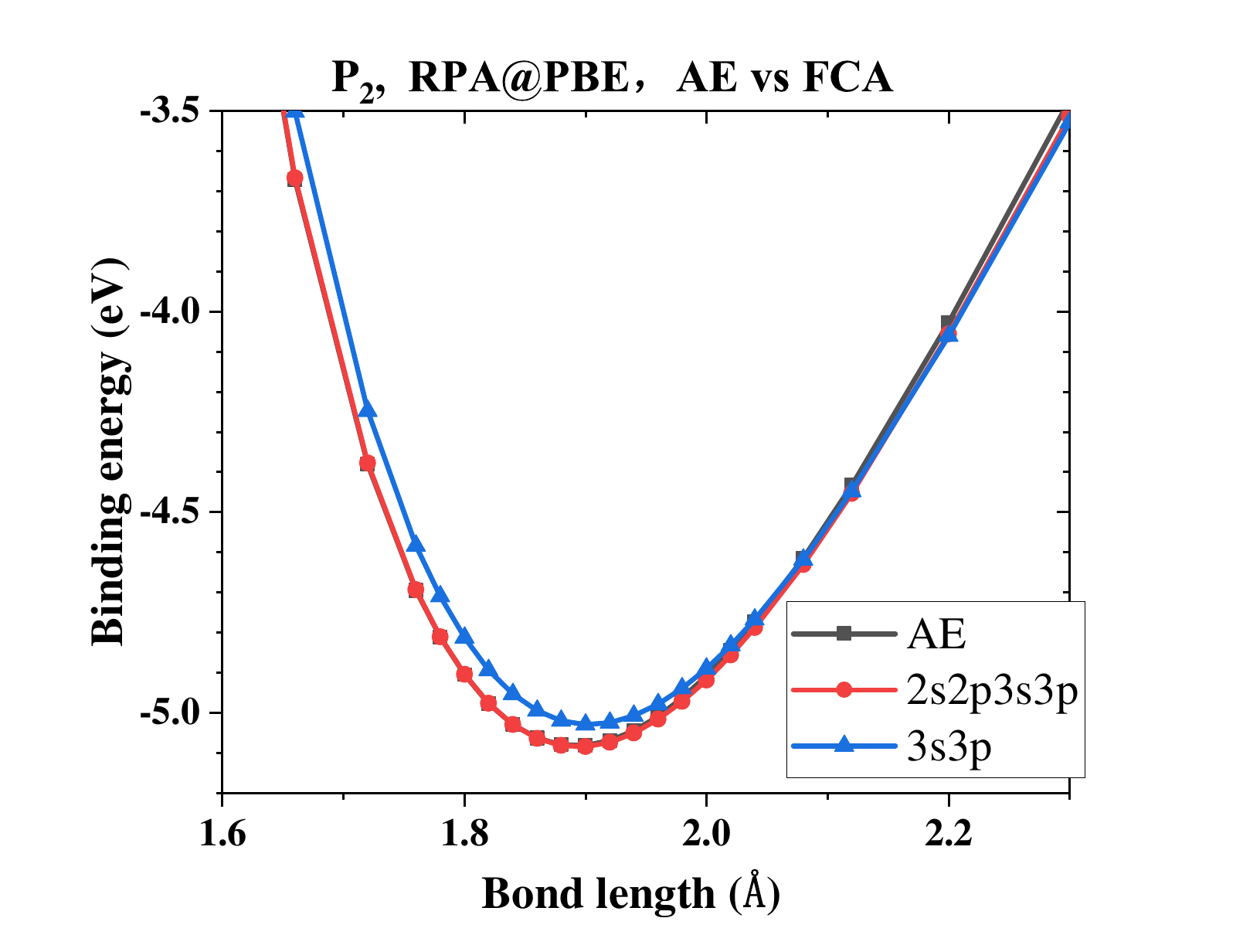}\\
     \includegraphics[scale=0.35]{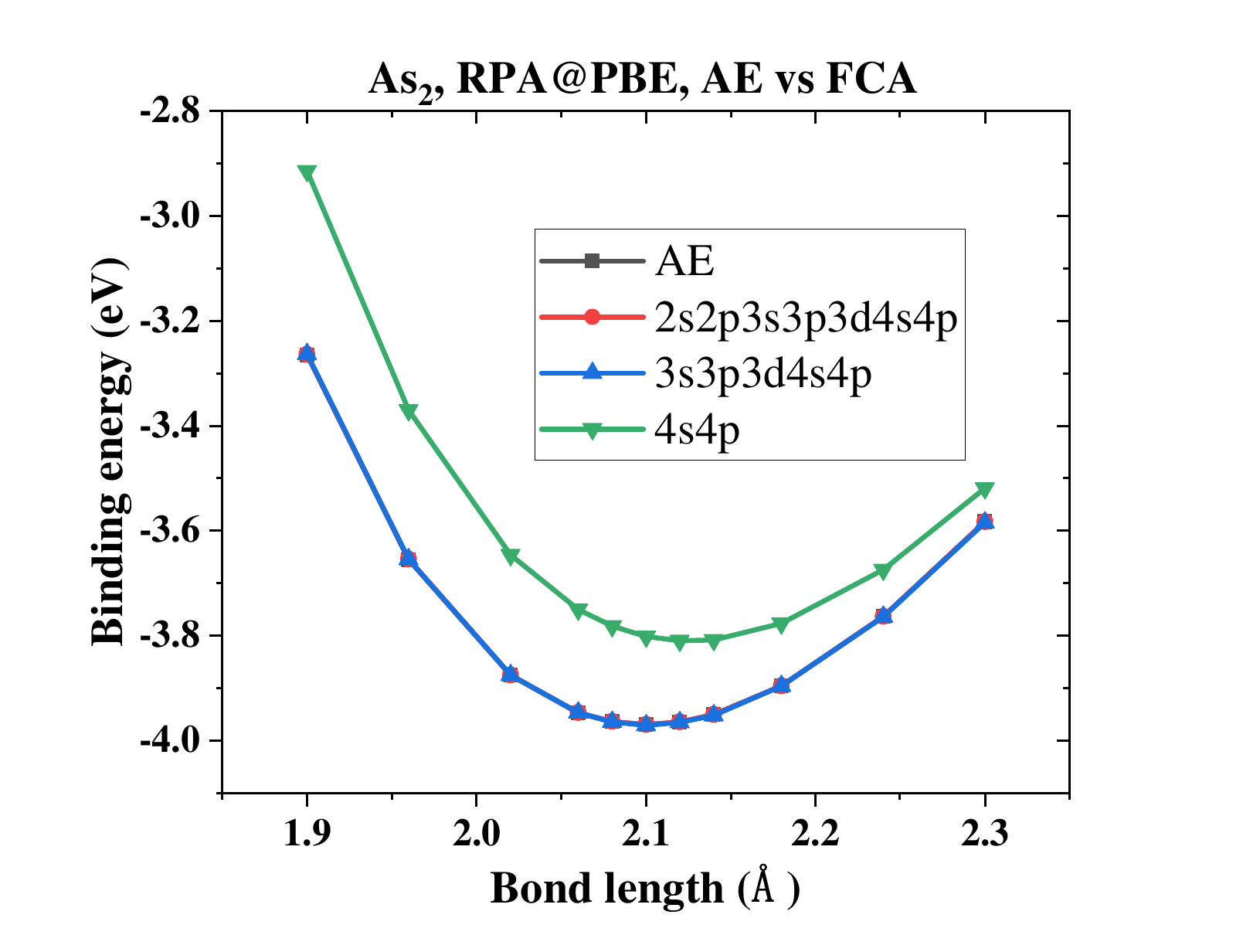}
    \caption{\label{fig:AE VS FCA} Binding energy curve of $\text{N}_2$,$\text{P}_2$,$\text{As}_2$ under AE and FCA. The notation of each curve denotes the electrons included in the calculation of the RPA correlation energy. For example, in the case of N$_2$, AE refers to all electrons, whereas $2s2p$ indicates that only the contributions to the RPA correlation energy from molecular orbital pairs formed by the $2s2p$ electrons are considered. }   
\end{figure}

The frozen-core approximation (FCA) is a common practice in correlated calculations. Under FCA, the electrons in the core are usually frozen at the level of DFT. The physical basis for doing so is that the core electrons do not directly participate in bonding, and so their contribution to the binding energy is much smaller than that of valence electrons. However, it is also not straightforward to give a precise error assessment of FCA. This is because in conventional approaches that rely on finite basis sets, fully converged AE and FCA results are difficult to obtain. However, the commonly used AO basis sets are not completely balanced in describing the valence electrons and core electrons; as such, the estimated FCA errors also depend on the basis sets used. In this work, we can provide both AE and FCA binding energies of diatomic molecules without BSIE, thereby offering an accurate assessment of the FCA error.

Here we perform AE and FCA calculations for molecules N$_2$, P$_2$, and As$_2$. In particular, for FCA calculations, different shells of core electrons are frozen, so that  we can monitor the effect of core electrons in each shell on the binding energy.  

We plot both the binding energy curves of AE and FCA of N$_2$, P$_2$, and As$_2$ in the upper, middle and lower panels of Fig.~\ref{fig:AE VS FCA}, respectively. 
For N$_2$, excluding the contribution from the core electrons ($1s$) results in a loss of binding energy of about 48 meV. For P$_2$, the innermost core electron ($1s$) has almost no contribution to the binding energy, while the contribution from the outer core shell ($2s2p$) is as large as 
52 meV.
For As$_2$, the innermost core electron ($1s$) and the second inner electron ($2s2p$) have almost no contribution to the binding energy, while the contribution from the outermost core shell
($3s3p3d$) is 160 meV. This investigation indicates that freezing all core electrons in FCA
calculations can result in non-negligible errors in the binding energies. This error becomes more
pronounced as the element gets heavier. However, by just including the outermost core shells in
the calculations, the results for P$_2$ and As$_2$ from FCA calculations are almost
indistinguishable from those from AE calculations, as can be seen from Fig.~\ref{fig:AE VS FCA}.

\section{FC-RPA binding energies for diatomic molecules}
In Table~\ref{tab:gw100} we present the FC-RPA@PBE binding energies of a set of 
diatomic molecules obtained both by the Sternheimer approach and by the traditional "sum over states" approach. The molecular set here is larger than those included in Table~III of the main
text and more representative.  The conventional "sum over states" results are obtained using both NAO (column 4) and GTO (columns 5 and 6) basis sets. Here we present FC-RPA results since
both NAO-VCC-$n$Z and cc-pV$n$Z basis sets are supposed to only describe the valence correlation.
The accurate dRPA-F12 results by Humer \textit{et al.} \cite{humer2022approaching} are again
included for comparison, but this time the FC approximation is used.

For NAO-based calculations (column 4), the CBS(4,5) results (extrapolation based on the NAO-VCC-4Z and 5Z basis sets) are provided. For GTO-based calculations, 
both the cc-pV6Z results (column 5) and the extrapolated CBS(5,6) results (column 6) in terms of cc-pV5Z and 6Z are presented. 
For some elements where cc-pV6Z is not available, the cc-pV5Z and CBS(4,5) results are presented instead. Those are marked with an asterisk. 
Furthermore, for the underlying PBE calculations, the relativistic effect is not considered for light elements with $Z\le20$, while for those with
$Z>20$, the atomic zeroth-order regular approximation (ZORA) \cite{Lenthe/Baerends/Snijders:1994}
as implemented in FHI-aims \cite{blum2009ab} is used. 

Table~\ref{tab:gw100} shows that the BSIE for the binding energies with the largest available NAO or GTO basis sets is still sizable. However, with the CBS extrapolation, the BSIE can be reduced to
within 0.4 kcal/mol.

\begin{table*}[!h]
\caption{RPA@PBE binding energy (in kcal/mol) for a set of diatomic molecules.  The third column represents the results obtained using the Sternheimer 
approach without basis set error. The fourth column (CBS-NAO) represents the extrapolated CBS results using NAO-VCC-4Z and 5Z. The fifth and sixth columns represent the results obtained using GTO cc-pV6Z and the extrapolated CBS(5,6) results based on cc-pV5Z and 6Z, respectively. For elements without cc-pV6Z basis set, the results in the fifth column are obtained using the cc-pV5Z, while the results in the sixth column are extrapolated from cc-pVQZ and cc-pV5Z. The results of these molecules are marked with a star. The last column lists very accurate results taken from the work of Humer \textit{et al} \cite{humer2022approaching}, obtained using explicitly correlated dRPA-F12 method. 
The mean absolute errors (MAE) are measured with respect to the reference results provided by the Sternheimer
method, averaged over the molecules where the results are available for other methods.}
\begin{tabular}{c c c c c c c}
\hline
\textbf{Molecule} & \textbf{Bond length  } & \textbf{This work}&\textbf{NAO[CBS(4,5)]} &\textbf{GTO[cc-pV6Z]} & \textbf{GTO[CBS(5,6)]} & \textbf{Ref.~\cite{humer2022approaching}} \\ \hline
H$_2$ & 0.74 & 108.72  & 108.72(0.00) &108.68(-0.04)& 108.77(0.05) & 108.69(-0.03) \\
He$_2$ & 4.30 &  0.001   & -0.001(-0.002) & 0.001(0.000) & 0.001(0.000) & / \\
Li$_2$& 2.70 & 18.84  &18.85(0.01) &18.50(-0.34)$^\ast$ &18.75(-0.09)$^\ast$ & 18.91(0.07)   \\
N$_2$ & 1.10 &  223.31 &   222.68(-0.63) & 221.50(-1.81)& 223.20(-0.11) & 223.34(0.03) \\
F$_2$ & 1.43 & 30.58  & 30.61(0.03) &29.69(-0.89) & 30.50(-0.08) & 30.56(-0.02)  \\
LiH & 1.60 &   54.65   &54.63(-0.02)&53.77(-0.88)$^\ast$ &54.39(-0.26)$^\ast$ &  54.48(-0.17)    \\
HF &  0.92     &  132.60  & 132.47(-0.13)&131.84(-0.76) &132.76(0.16) &   132.59(-0.01) \\
LiF & 1.58  & 127.25 & 127.02(-0.23) &124.20(-3.05)$^\ast$& 126.64(-0.61)$^\ast$ & 127.20(-0.05) \\
CO  &1.14  &  244.47 &244.07(-0.40)&242.95(-1.52) & 244.26(-0.21) &244.46(-0.01) \\
Ne$_2$& 3.24 & 0.041 &-0.006(-0.047) &0.025(-0.016) &0.030(-0.011) & / \\
Na$_2$ & 3.18 & 13.24 &13.04(-0.20) &12.93(-0.31)$^\ast$&13.26(0.02)$^\ast$ & / \\
K$_2$ & 4.14 & 9.06  &/&/& / &/\\
Rb$_2$ & 4.50 & 8.21 &/&/& /&/ \\
P$_2$ & 1.91  &  115.96& 115.18(-0.78)&113.62(-2.34)&115.13(-0.83)&/ \\ 
As$_2$ & 2.06 & 86.48&/&83.98(-2.50)$^\ast$&85.96(-0.52)$^\ast$ &/ \\
Cl$_2$ & 2.02 &  49.76 &48.92(-0.84)&48.32(-1.44)&49.57(-0.19)& / \\
Ar$_2$ & 3.84 & 0.200 &0.077(-0.123)&0.182(-0.018) &0.199(-0.001) &/ \\
Br$_2$  & 2.24& 40.67 &/&38.32(-2.35)$^\ast$&39.99(-0.68)$^\ast$& / \\
I$_2$  &  2.65 & 84.72     &/&/&/ &/\\
KH   &   2.34 & 41.33  & / &/&/&/\\
HCl  & 1.28 & 100.55 & 100.38(-0.17)&99.76(-0.79)&100.45(-0.10)&/ \\
BF   & 1.27  & 167.87 &167.86(-0.01)&166.78(-1.09) &167.78(-0.09)&/ \\
BrK  & 2.77 & 80.40 & / &/&/&/\\
NaCl & 2.40 & 90.59 &89.73(-0.86)&87.46(-3.13)$^\ast$&89.50(-1.09)$^\ast$& / \\
BN &1.30 &102.96 & 102.80(-0.16)&101.53(-1.43)&102.91(-0.05)&/ \\
PN & 1.50 & 144.47 & 144.08(-0.39)&142.47(-2.00)&143.87(-0.60)&/ \\
BeO & 1.35 & 100.94&100.72(-0.22)&99.10(-1.84)&100.76(-0.18) & / \\
MgO & 1.78 & 60.29 &59.86(-0.43)&57.09(-3.20)$^\ast$&59.13(-1.16)$^\ast$& / \\
Kr$_2$ & 4.10 & 0.291 & 0.260(-0.031) & 0.248(-0.043)& 0.290(-0.001) &/ \\
Ag$_2$ & 2.53 & 33.65 & /&/&/&/ \\
Zu0  &   1.72 &   34.57 & / &32.30(-2.27)$^\ast$ &34.08(-0.49)$^\ast$ &/ \\ 
\hline  \hline
MAE    &    \    & /
&   0.32  &1.60     &  0.36  & 0.05 \\  
      \hline
      \hline
     \label{tab:gw100}
\end{tabular}
\end{table*}

\section{Evaluation of basis-set-error for SOS-MP2 method}
The essential point of our work is that we are able to eliminate the BSIE of the density response function. Thus, in addition to high-precision RPA correlation energy calculation, our method can also be used for correlation methods such as GW and opposite spin (OS) MP2\cite{jung2004scaled}, etc. Here, we show how to accurately evaluate the BSIE of the OS-MP2 method using our techniques. It should be pointed out that our current work is based on the ground state of KS-DFT, while the standard MP2 method requires the Hartree-Fock orbitals as the reference state. For convenience,
in this test we used the KS-DFT as the reference point to calculate the MP2 correlation energy. 
Our focus here is to evaluate the BSIE of the method rather than the absolute results of the calculation. 
Extending this technique to the Hartree-Fock ground state is straightforward.

Jung et al. proposed scaled-opposite-spin  MP2 (SOS-MP2) method \cite{jung2004scaled} based on the spin-component-scaled  MP2 (SCS-MP2) method \cite{grimme2003improved} proposed by Grimme. Simply speaking, the standard MP2 correlation energy can be divided into the same spin (SS) part and the opposite spin (OS) part. 
\begin{equation}
    E_\text{MP2}=E^\text{OS}_\text{MP2}+E^\text{SS}_\text{MP2}
\end{equation}
Grimme pointed that multiplying the two parts by different scaling coefficients can provide a better description of the ground state energy of the molecule compared to the standard MP2 method. This method is called SCS-MP2,
\begin{equation}
    E_\text{SCS-MP2} = C_\text{OS}E^\text{OS}_\text{MP2} + C_\text{SS}E^\text{SS}_\text{MP2} \, 
\end{equation}
where $C_\text{OS}=\frac{6}{5}$ and $C_\text{SS}=\frac{1}{3}$ is a suitable choice. On this basis, Jung et al. proposed that only calculating the OS part of the MP2 correlation energy  and scaling it lead to comparable results with the SCS-MP2 method,
\begin{equation}
    E_\text{SOS-MP2} = C_\text{OS}E^\text{OS}_\text{MP2} 
\end{equation}
where $C_\text{OS}$ is suggested to be $1.3$. SOS-MP2 is very advantageous because the computationally
involved part, the exchange contribution to the SS correlation, does not need to be handled. 
That is, the SOS-MP2 method not only improves the accuracy of MP2, but also reduces the computational complexity. Moreover, since the OS part of MP2 correlation energy can be obtained through the density response function,
\begin{equation}
    E_\text{MP2}^\text{OS} = -\int_{0}^{\infty} \frac{\mathrm{d} \omega}{4 \pi} \sum_{\substack{\sigma=\alpha, \beta \\ \gamma \neq \sigma}} \operatorname{Tr}\left[v \chi_{0}^{\gamma} v \chi_{0}^{\sigma}\right]\, 
\end{equation}
high-precision OS-MP2 (SOS-MP2) correlation energy without BSIE can be obtained using the technique developed in the present work.

\begin{figure}[htbp]
   \centering
    \includegraphics[scale=0.3]{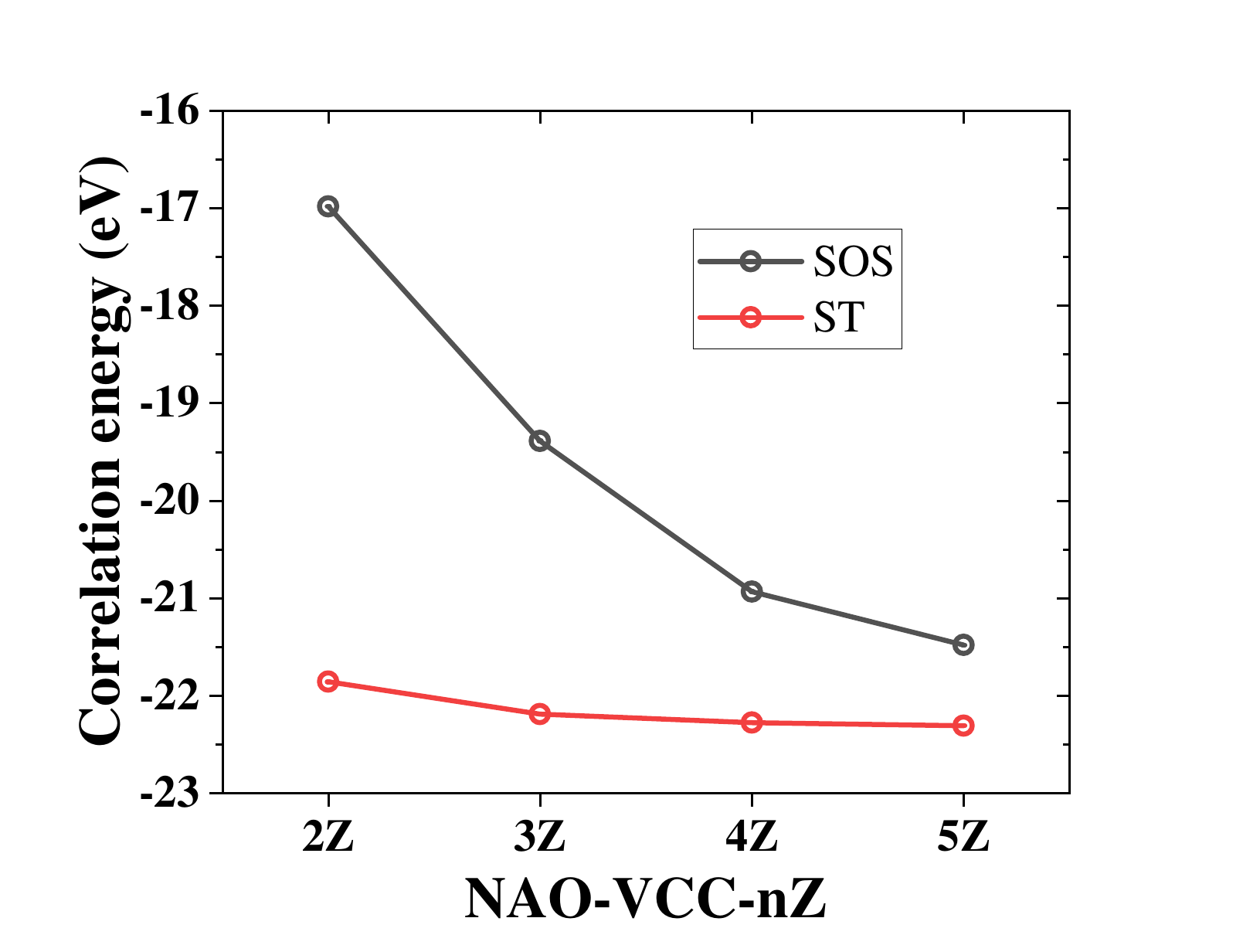}
    \includegraphics[scale=0.3]{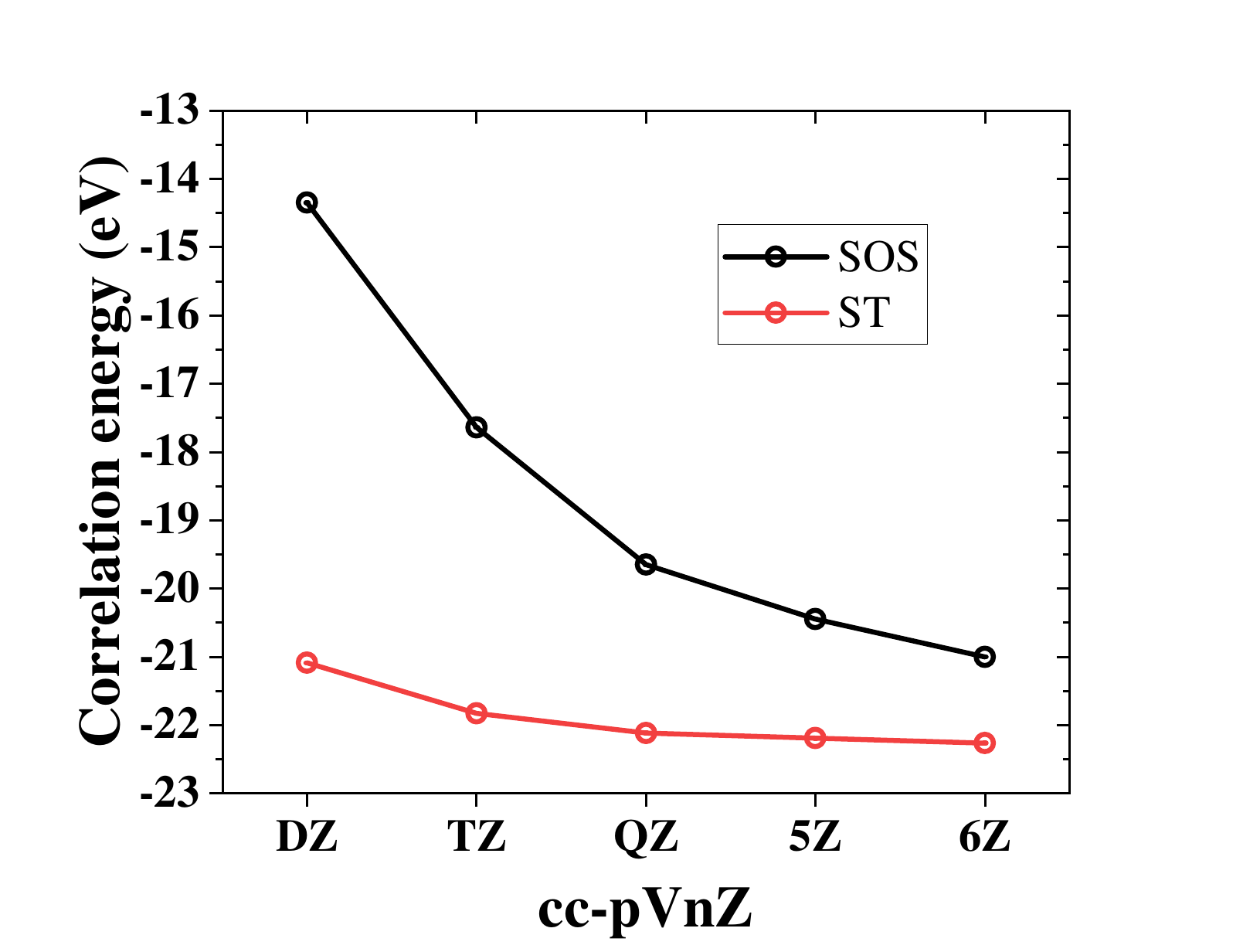}\\ 
    \caption{\label{fig:OS-MP2} SOS-MP2 correlation energy  of N$_2$ under NAO-VCC-nZ and cc-pVnZ basis sets. {In the curve notation, SOS denotes results obtained using the sum-over-states approach, while ST denotes those obtained with the Sternheimer method.}}
\end{figure}

In Fig.~\ref{fig:OS-MP2}, we demonstrate the convergence behavior of the SOS-MP2 correlation energy with respect
to the basis size for the N$_2$ molecule. The results obtained using both NAO-VCC-$n$Z and cc-pV$n$Z basis sets
are presented. Here we see that the SOS-MP2 energy within the Sternheimer approach converges quickly to the CBS limit, while the traditional "sum over states" approach converges much slower, ending up with
sizable BSIE even with the largest available basis set in each series. We note that the slight change of the
SOS-MP2 correlation energy with the basis size comes from the fact that we are using the RI
approach to MP2, where the results depend on the auxiliary basis set. In our implementation, the auxiliary basis 
functions are generated on the fly based on the single-particle basis set, and hence the mild dependence. 
In summary, the convergence behavior observed for the RI-SOS-MP2 energy is the same as the RI-RPA correlation
energy, as demonstrated in Ref.~\onlinecite{peng2023basis}. 

\section{Numerically precise all-electron RPA correlation energies of General molecules based on finite element method  }
\begin{figure}[htbp]
   \centering
    \includegraphics[scale=0.35]{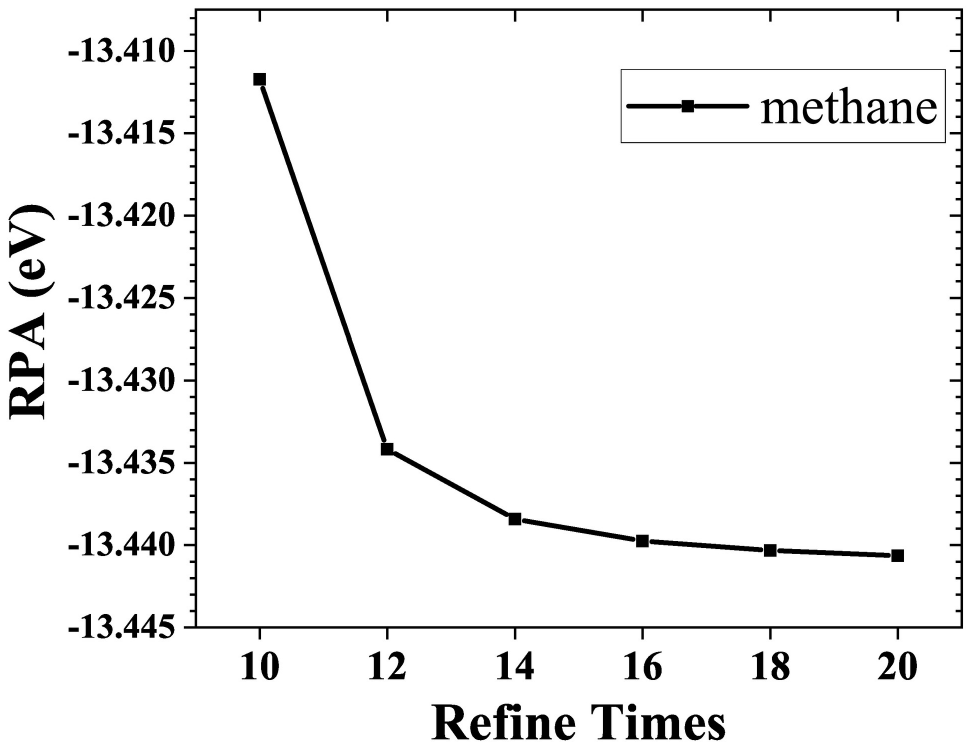}
    \caption{\label{fig:CH4_FEM} The RPA correlation energy of the methane molecule as a function of finite element grid density  } 
\end{figure}

For general polyatomic molecules, the prolate spheroidal coordinate system is not applicable. However, this doesn't  mean that our general methodology is limited to diatomic molecules. In fact, we have already obtained some results for polyatomic molecules. By utilizing adaptive refinement techniques in the finite element space, we can converge the RI-RPA correlation energy to the meV level.
{The finite element method (FEM) is a real-space discretization approach that offers greater flexibility  compared to finite difference methods (FDM).}
Additionally, adaptive refinement techniques can compute the error for each element and selectively refine the elements with the largest errors, thereby generating efficient grids for describing general molecules. Below, we present the results for the methane molecule as an example.
Here, we calculate the methane molecule at its equilibrium structure, with the C atom located at the origin. The auxiliary basis set is generated "on-the-fly" by FHI-aims using NAO-VCC-3Z basis set. The methane molecule is placed inside a cubic box of  $30\times 30 \times 30$~Bohr$^3$ 
and the Sternheimer equation is solved in the finite element space with adaptive refinement. The finite element software we use here is the open-source OpenPFEM \cite{Liao2025OpenPFEM}, source code is available from \url{https://gitlab.com/xiegroup/OpenPFEM1.0_public}. The relationship between the grid density and the RPA correlation energy for the methane molecule is shown in Table~\ref{tab:methane}.
\begin{table*}[!h]
\caption{Convergence behavior of the RPA correlation energy of the methane molecule with adaptive refinement in the finite element space. The first column represents the number of adaptive refinements, the second column represents the total number of degrees of freedom (DOF), and the third column represents the RPA correlation energy. The last row provides the result obtained using the traditional ``sum-over-states" method.}
\begin{tabular}{c c c }
\hline
\textbf{Refine } & \textbf{DOF  } & \textbf{RPA (eV)} \\ \hline
10 & 73183 & -13.41173   \\
12 &106317 &-13.43417\\
14 &122859&-13.43841\\
16 &151961 &-13.43975\\
18& 189719&-13.44033\\
20&225073&-13.44064\\
SOS&/&-11.65628\\
\hline  \hline

      \hline
      \hline
     \label{tab:methane}
\end{tabular}
\end{table*}
As can be seen, the RPA correlation energy gradually converges with increasing grid refinement. From 18 to 20 refinements, the absolute energy changes by only 0.3 meV. It can also be seen that the ``sum over states'' result obtained with the NAO-VCC-3Z single-particle basis set covers approximately 87$\%$ of the correlation energy. Figure~\ref{fig:CH4_FEM} shows the convergence curve of the RPA correlation energy with respect to the number of adaptive grid refinements. From the trend, the RPA correlation energy converges to within 1 meV when the number of grid degrees of freedom reaches 220,000. In the prolate spheroidal coordinate system, the number of grid points we use is on the order of tens of thousands. Therefore,the degrees of freedom in the finite element space increase by only one order of magnitude, which is acceptable.

\begin{figure}[htbp]
   \centering
    \includegraphics[scale=0.10]{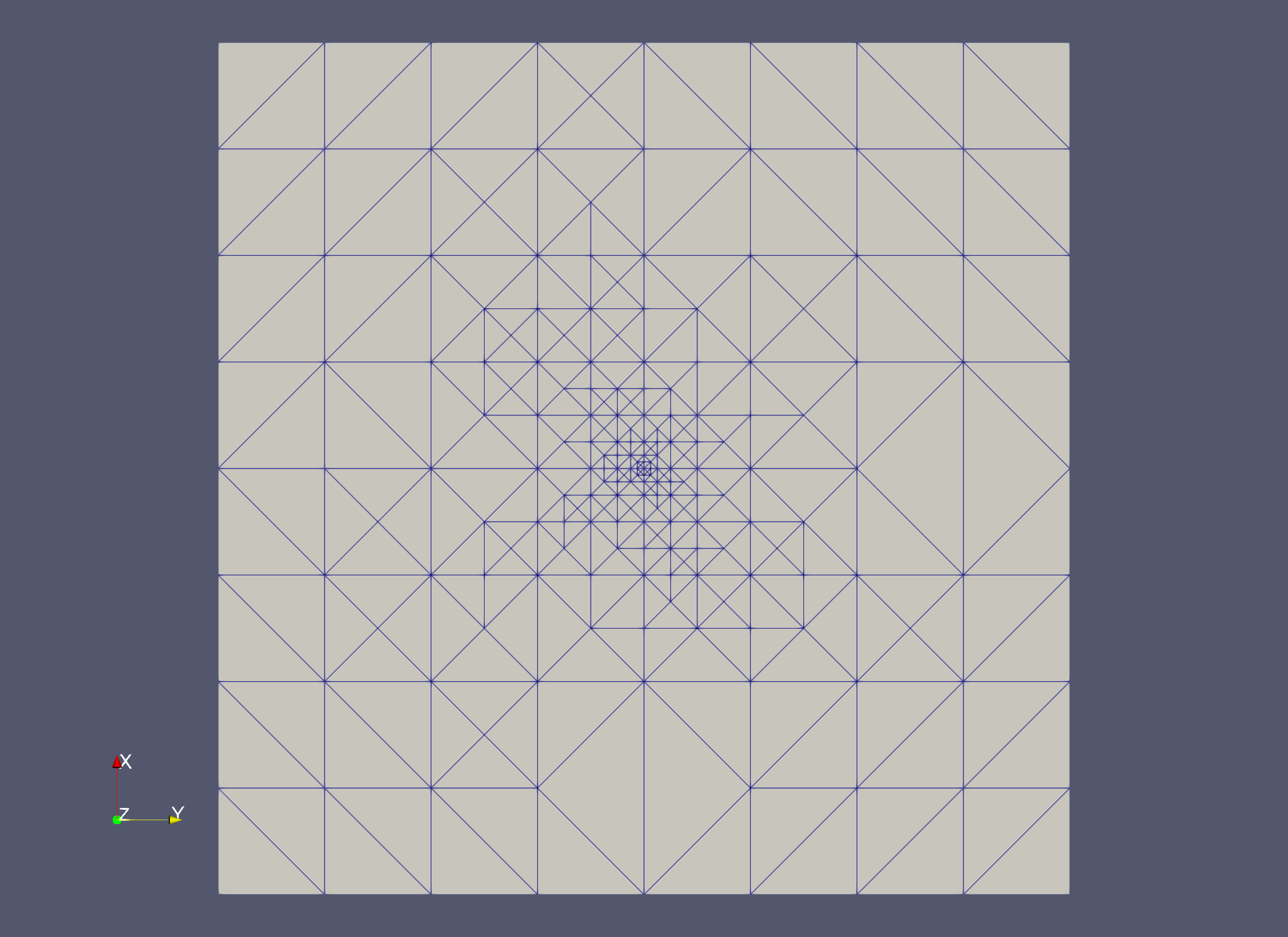}
    \includegraphics[scale=0.10]{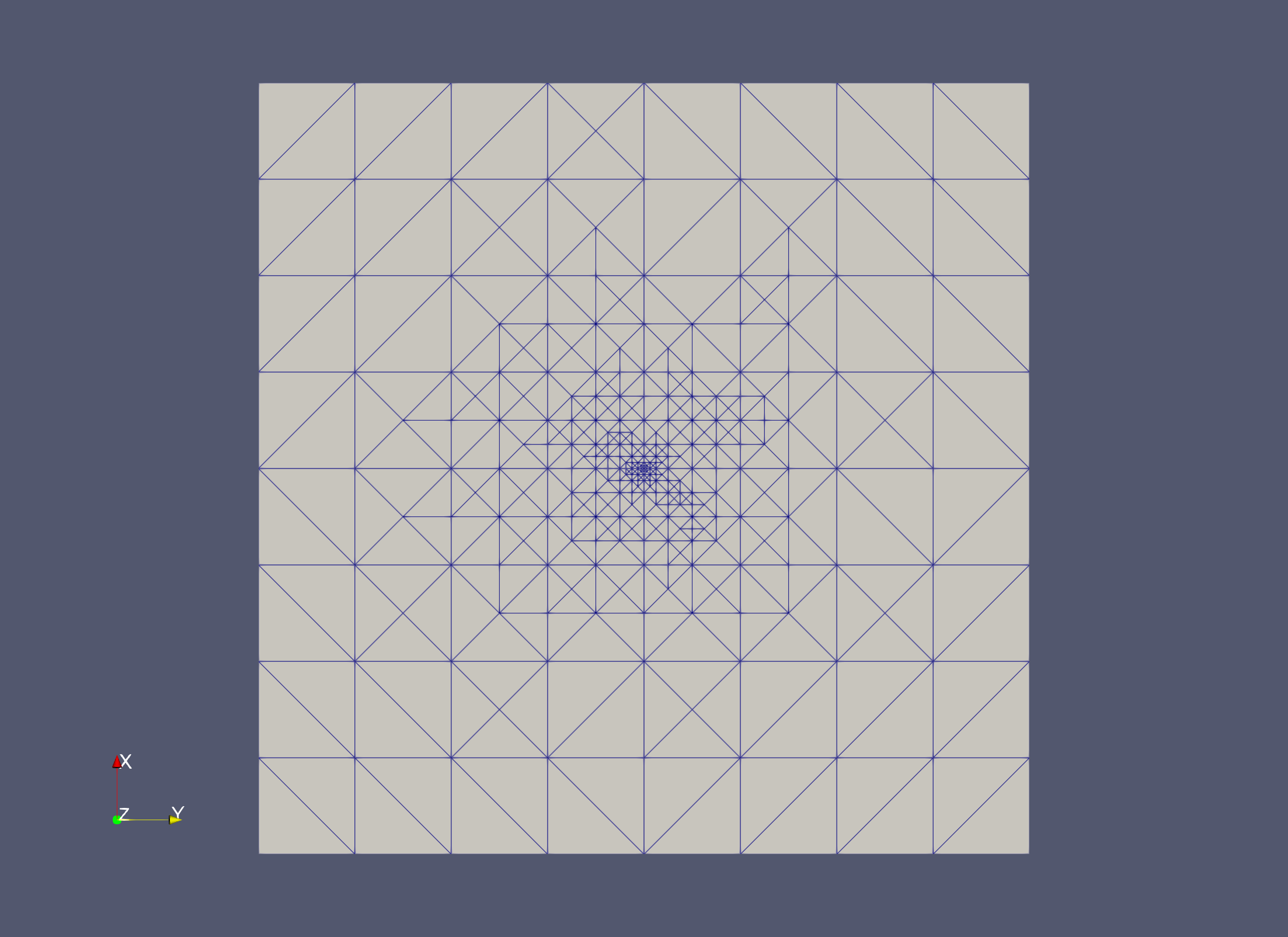}\\ 
        \caption{\label{fig:FEM_grid} Schematic diagram of finite element grid points. A cross-sectional view of the finite element mesh space along the X-Y plane passing through the origin is shown below. The upper image represents the result after 12 adaptive refinements, while the lower image shows the result after 20 adaptive refinements. The center of the image corresponds to the origin, which is the position of the C atom.} 
\end{figure}

In Fig.~\ref{fig:FEM_grid}, we show the cross-sectional view of the finite element space in the X-Y plane. It can be seen that the finite element mesh is densest near the C atom (the origin), while the mesh is sparsest in the boundary regions. Figure~S8 also illustrates the refinement process of the finite element mesh. From 12 to 20 refinements, the mesh near the atomic nuclei becomes very dense. This is the foundation that allows us to perform all-electron calculations.

Finally, it should be noted that iterative diagonalization can also be used in the finite element space to eliminate errors from the auxiliary basis set. In our future work, we will further assess the feasibility of this approach.

\newpage
\bibliography{RPA_bib}

\end{document}